\documentclass{aa}  

\usepackage{multirow}  
\usepackage{booktabs}  
\usepackage{graphicx}
\usepackage{txfonts}
\usepackage{hyperref}
\usepackage{orcidlink}
\usepackage{color}
\usepackage{natbib}
\usepackage{ulem}
\usepackage{amsmath,amssymb}
\usepackage{soul} 

\bibpunct{(}{)}{;}{a}{}{,} 

\newcommand{\msun}{\ensuremath{\mathrm{M}_\odot}}

\newcommand{\nbo}{\textsc{Nbody6++GPU}}
\newcommand{\nbody}{\textit{N}-body}

\begin{document} 

\title{Simulating tidal disruption events in nuclear star clusters}
\titlerunning{Simulating tidal disruption events in nuclear star clusters}

\author{
Philip Cho\inst{1}\thanks{E-mail: philip.cho@stud.uni-heidelberg.de}
\and
Kai Wu\orcidlink{0000-0003-0349-0079}\inst{1}
\and
Francesco Flammini Dotti\orcidlink{0000-0002-8881-3078}\inst{2,3,4}
\and
Taras Panamarev\orcidlink{0000-0002-1090-4463}\inst{5}
\and
Shuo Li\orcidlink{0000-0001-6530-0424}\inst{6}
\and
Shiyan Zhong\orcidlink{0000-0003-4121-5684}\inst{7}
\and 
Klaus Reuter\inst{8}
\and
Rainer Spurzem\orcidlink{0000-0003-2264-7203}\inst{1,9,10}
}
\authorrunning{Cho et al.}
\institute{
Astronomisches Rechen-Institut, Zentrum f\"ur Astronomie der Universit\"at Heidelberg, M\"onchhofstr. 12-14, D-69120 Heidelberg, Germany
\and Department of Physics, New York University Abu Dhabi, PO Box 129188 Abu Dhabi, UAE
\and Center for Astrophysics and Space Science (CASS), New York University Abu Dhabi, PO Box 129188, Abu Dhabi, UAE
\and Dipartimento di Fisica, Sapienza, Universit\'a di Roma, P.le Aldo Moro, 5, 00185 - Rome, Italy
\and Fesenkov Astrophysical Institute, Observatory 23, 050020, Almaty, Kazakhstan
\and National Astronomical Observatories, Chinese Academy of Sciences, Beijing 100012, People’s Republic of China
\and Southwestern Institute for Astronomy Research, Yunnan University, Chenggong District, Kunming 650500, Yunnan, China
\and Max Planck Computing and Data Facility, Giessenbachstrasse 2, D-85748 Garching, Germany
\and National Astronomical Observatories, Chinese Academy of Sciences, 20A Datun Rd.,Chaoyang District, 100101, Beijing, China
\and Kavli Institute for Astronomy and Astrophysics, Peking University, Yiheyuan Lu 5, Haidian Qu, 100871, Beijing, China
}

\date{Received XXX; accepted YYY}

\abstract
   {Nuclear star clusters (NSCs) and supermassive black holes (SMBHs) often coexist at the centers of galaxies, engaging in complex dynamical interactions that influence SMBH growth. Stars that approach the SMBH too closely are tidally disrupted, and a fraction of their debris is accreted onto the black hole. Such events may contribute to SMBH growth in gas-deficient galaxies.}
   {Our work aims to improve the theoretical framework for understanding the mass accretion of stellar debris following tidal disruption events (TDEs) through numerical simulations. We perform direct $N$-body simulations of an NSC with a central SMBH embedded in a fully self-consistent dynamical model to test a new TDE accretion prescription in a set of pilot simulations. }
   {The previous TDE prescription used a fixed accretion radius and assumed that the entire stellar mass was accreted after disruption. In the new partial-mass accretion model, the mass added to the SMBH depends on the orbital eccentricity and penetration factor of the disrupted star.}
    {We find that most TDEs involving main-sequence stars are marginally eccentric or marginally hyperbolic and contribute approximately half of the stellar mass to SMBH growth, consistent with recent hydrodynamical simulations. In all simulated models, the TDE-driven mass accretion rate exhibits an early peak followed by a rapid decline. Models with higher initial SMBH masses exhibit a faster decline in the accretion rate, consistent with more rapid loss-cone depletion and inefficient replenishment.}
   {}


\maketitle

\section{Introduction} \label{sec:intro}
Supermassive black holes (SMBHs) are commonly found at the centers of galaxies whose stellar masses range from $10^{10}$ to $10^{12}\,\mathrm{M_{\odot}}$ \citep{Kormendy_1995,Vika_2009,Kormendy_2013}. Galaxies below this mass range often host nuclear star clusters (NSCs) at their centers, with such clusters observed in galaxies with stellar masses as low as $\sim10^6\,\mathrm{M_{\odot}}$ \citep{Neumayer_2020}. These extremely dense stellar systems are highly luminous and may constitute the dominant nuclear component in low-mass galaxies \citep{Ferrarese_2006}. However, NSCs and massive black holes coexist in some galaxies \citep{Filippenko_2003, Graham_2009}. The best-known example is the Milky Way, which hosts an inactive SMBH of $\sim 4\times10^6 \mathrm{M_{\odot}}$ within an NSC of $\sim 10^7 \mathrm{M_{\odot}}$ \citep{genzel10}. In such a dense environment, stars and compact objects undergo dynamical processes that may bring them close to the SMBH, resulting in violent astrophysical phenomena such as stellar tidal disruption events (TDEs).

When a star on a low-angular-momentum orbit passes sufficiently close to an SMBH, the tidal force exerted by the SMBH can exceed the self-gravity of the star and destroy it. The characteristic distance at which this occurs is called the tidal radius. The associated flare is generally attributed to the fallback, circularization, and subsequent accretion of stellar debris, through which gravitational binding energy is converted into radiation, potentially exceeding the Eddington luminosity \citep{Hills1975, rees88, enk89}. When the star is completely destroyed in a full TDE (FTDE), roughly half of the stellar material becomes gravitationally bound and falls toward the SMBH, while the remaining material gains energy through tidal interactions and escapes the SMBH's potential well \citep{rees88}. If the dense stellar core survives and only the outer layers are removed, the encounter is classified as a partial tidal disruption event (PTDE), which generally produces a fainter light curve than an FTDE \citep{Gomez.et.al_2020,MacLeod2012,Bogdanovic2014,Wang_2026}.

Although TDEs are intrinsically rare, their flares have been detected across multiple wavelengths. The first candidates were identified in the early 1990s through soft X-ray emission in the 0.1--2.4 keV band from otherwise quiescent galaxies in the ROSAT All-Sky Survey (RASS) \citep{Bade_1996, komossa99, Donley_02}. Additional candidates not detected by RASS were later identified in the XMM-Newton Slew Survey, which covers $\sim 15\%$ of the sky in the 0.2--2 keV band \citep{Esquej_07}. In the ultraviolet, the GALEX Deep Imaging Survey discovered three TDE candidates in inactive galaxies, two of which were also observed in the \textit{g}, \textit{r}, \textit{i}, and \textit{z} optical bands by the CFHT Legacy Survey (CFHTLS) \citep{Gezari_2006, gezari08}. The advent of high-cadence, wide-field optical surveys has significantly increased the number of TDE detections and improved the temporal sampling of their light curves.

To date, roughly two hundred TDE candidates have been identified as transients in galactic nuclei, with emission detected from radio to gamma-ray wavelengths \citep[][and references therein]{Komossa_Bade_1999, Komossa_2015, Gezari2021}. A recent study based on the first two eROSITA all-sky surveys identified 31 X-ray-selected TDE candidates and found that their X-ray luminosity function and volumetric rate are consistent with previous observational and theoretical expectations \citep{eROSITA_2025}. A substantial fraction of these events also exhibit multi-wavelength flares in the optical, mid-infrared, and radio bands. \citet{Franz_2026} presented the Open mulTiwavelength Transient Event Repository (OTTER), a publicly available catalog and analysis framework containing more than 118,000 observations of 240 TDE candidates from radio to X-ray wavelengths. It provides the largest multi-wavelength photometric archive of TDEs to date. A small number of observed TDE candidates have been associated with binary SMBH systems, in which the light curves may differ from those of TDEs around single SMBHs \citep{Liu_2014, Shu_2020, Huang_2021}. In such systems, perturbations from the companion SMBH can produce repeated gaps or truncated features in the light curve \citep{Liu_2009, Ricarte_2016, Coughlin_2017, Vigneron_2018}.

Wide-field missions such as Einstein Probe \citep{Yuan_2025} and the forthcoming ULTRASAT \citep{ULTRASAT_2024}, together with surveys such as LSST \citep{LSST_2019, Bricman_2020}, are expected to increase the number of observed TDEs, with LSST alone predicted to discover thousands annually. ELT and GMT will enable detailed optical and near-infrared characterization of these events, while SKA will provide complementary radio observations of their jets and outflows \citep{Wheeler2019ELT, GMTScienceBook2018, donnarumma2015skapowerfulhunterjetted}. TDEs at high redshift may also become detectable through gravitational lensing by intervening galaxies because such lensing can increase their apparent brightness. Such observations would help constrain TDE rates at earlier cosmic epochs \citep{Szekerczes_2024, Chen_2024}. Numerical studies have also progressed considerably in recent years. Hydrodynamical simulations model the disruption process and the resulting debris fallback rates, which provide an important physical basis for interpreting observed TDE light curves \citep{Guillochon_2013, Goicovic2019, Law-Smith2020, lodato_2020, Ryu_2020, Liu_2025}. These studies also show that only a fraction of the disrupted stellar material remains gravitationally bound and contributes to mass accretion.

Numerical simulations provide a means of investigating the coupled dynamical evolution of NSCs and their central SMBHs. TDEs can constrain SMBH demographics and contribute to SMBH growth in gas-poor galactic nuclei, where the supply of gas available for accretion is limited \citep{stone_rates_2020}. Early work suggested that TDEs could contribute substantially to the growth of SMBHs with masses $\lesssim 2\times10^6\,\msun$ \citep{Milo2006}. The supply of stars available for TDEs is governed by loss-cone dynamics. The loss-cone comprises stellar orbits with sufficiently low angular momentum that their pericenter distances satisfy $r_{\rm p}\leq r_{\rm t}$ \citep{Merritt_2013}. Stars on these orbits are susceptible to tidal disruption by the SMBH. After the initially populated loss-cone is depleted, two-body relaxation can scatter other stars onto loss-cone orbits. \citet{Zhong_2014} investigated TDE rates and loss-cone dynamics with direct $N$-body simulations and confirmed that most tidally disrupted stars originate near the critical radius. \citet{panamarev2019} extended these calculations to one million particles and followed the evolution of the Galactic Center and its NSC. They obtained a TDE rate consistent with theoretical expectations and observational estimates for Milky Way-like galaxies. \citet{Rizzuto_2023} shows that tidal captures and TDEs can also contribute to the growth of intermediate-mass black holes.

However, these previous $N$-body studies assumed a simplified treatment of mass accretion following a TDE, in which the entire stellar mass was removed from the system and added to the black hole. This assumption does not capture the results of hydrodynamical simulations, which show that only part of the stellar debris remains gravitationally bound, while the remainder becomes unbound. Consequently, previous models may have systematically overestimated SMBH growth. Over time, this overestimate can affect both the tidal disruption rate and the subsequent evolution of the NSC.

A realistic description of SMBH growth requires an accretion model that distinguishes between bound and unbound debris following a TDE, with the mass added to the SMBH depending on the orbital properties of the disrupted star. Such a partial-mass accretion (PMA) model provides a more consistent framework for assessing the cumulative contribution of TDEs to long-term SMBH growth. In this work, we implemented such a prescription in the direct \nbody{} code \nbo{} and tested it using simulations of NSCs hosting central SMBHs. We find that most TDEs result in the accretion of only a fraction of the stellar debris, with approximately half of the disrupted star's mass added to the SMBH. The new treatment systematically reduces the SMBH growth rate compared with the previous full-mass accretion (FMA) model.

This paper is organized as follows. In \hyperref[sec:method]{Section~\ref*{sec:method}}, we describe the improved TDE accretion prescription implemented in the direct $N$-body code \textsc{Nbody6++GPU} and the suite of simulations used to test it. In \hyperref[sec:results]{Section~\ref*{sec:results}}, we present the resulting SMBH growth histories, the dynamical response of the NSC, and the statistical properties of the TDEs. In \hyperref[sec:emri]{Section~\ref*{sec:emri}}, we discuss limitations of the current treatment of compact object accretion and summarize an approximate EMRI-motivated capture criterion. Finally, in \hyperref[conclusion]{Section~\ref*{conclusion}}, we summarize our conclusions and outline directions for future work.

\section{Methodology} \label{sec:method}
Interpreting the rapidly growing samples of TDEs requires self-consistent dynamical models that connect individual stellar encounters with the long-term evolution of SMBHs in NSCs. We therefore used the direct \nbody{} code \nbo{}, which provides a high-accuracy method for modeling collisional stellar systems.
\subsection{\nbo}
This study employs the high-resolution code \nbo. The code is optimized for high-performance computing using MPI, SIMD, OpenMP, and GPU parallelization \citep[see][and references therein]{spurzem1999direct, NitadoriAarseth2012, Wangetal2015, Spurzem_2023}. \nbo{} belongs to the family of direct \nbody{} integration codes developed from the pioneering work of Sverre Aarseth \citep[see][and references therein]{Aarseth1985, spurzem1999direct, Aarseth1999b, Aarseth1999a, Aarseth2003, AarsethEtal2008}. The code can resolve close binaries and higher-order stellar subsystems, allowing detailed simulations of realistic star clusters. It achieves this precision through several numerical techniques, including the Hermite integration scheme with hierarchical block time steps \citep{McMillan1986, HutEtal1995, Makino1991, Makino1999}, Kustaanheimo--Stiefel (KS) regularization \citep{kustaanheimo1965}, and the Ahmad--Cohen (AC) neighbor scheme \citep{AhmadCohen1973}. The evolution of single and binary stars follows the prescriptions of \citet{Hurley_2000, Hurley_2002, Church_Tout_Hurley_2009, Kamlah2022, Spurzem_2023}.

We used a special variant of \nbo{}, called \textsc{Stardisk}, to model an NSC-like stellar system hosting a central massive object that represents an SMBH \citep{CHO_IAU}. This variant incorporates the algorithms described above and introduces additional features designed for galactic nuclei. The features relevant to the present study are as follows:
\begin{itemize}
    \item The SMBH is treated as a fixed potential at the cluster center, which is a reasonable approximation given the negligible motion of SMBHs in real NSCs \citep{Lin1980}.
    \item Because the SMBH is not represented by an \nbody{} particle, it is excluded from KS regularization.
    \item The code includes an option to model a gaseous accretion disk for simulations of active galactic nuclei \citep{Panamarev2018}. However, we did not use this module in the present study.
\end{itemize}
The \textsc{Stardisk} framework has been developed and applied in several previous studies of TDEs and compact object accretion \citep{Just2012, Kennedy2016, panamarev2019}. Both the standard \nbo{} code\footnote{\url{https://github.com/nbody6ppgpu}} and its \textsc{Stardisk} variant\footnote{\url{https://github.com/nbody6ppgpu/Nbody6PPGPU-beijing/tree/stardisk-dev}} are publicly available on GitHub. In \autoref{appendix:profiling} we show for future applications how well the code scales up to more than ten million particles.

The primary aim of this work is to examine how a revised TDE accretion prescription affects SMBH growth and the evolution of the surrounding NSC. In the FMA prescription adopted by \citet{panamarev2019}, mass accretion was modeled using a fixed numerical accretion radius. Any star that passed within this radius was removed from the system, and its entire mass was assumed to be accreted by the SMBH. As discussed in \hyperref[sec:intro]{Section~\ref*{sec:intro}}, this assumption is not always physically accurate because only a fraction of the disrupted stellar debris remains gravitationally bound. In the present work, we replace the FMA prescription with a PMA prescription based on the bound debris fraction.

Apart from the accretion prescriptions, the two models use identical initial conditions. Because the prescriptions differ in both the accreted debris fraction and the evolution of the numerical accretion radius, our comparison examines their combined effect on SMBH growth and NSC evolution. The revised accretion treatment is described in detail in the following subsections.

\subsection{Stellar TDEs and Mass Accretion} \label{subsec:2.2}
A star is considered tidally disrupted when its orbital pericenter distance from the SMBH is smaller than the tidal radius $r_{\rm t}$, which is given by
\begin{equation} \label{eqn:tidal}
    r_{\rm t} = r_{\star}\left(\frac{M_{\rm SMBH}}{m_{\star}}\right)^{1/3},
\end{equation} 
where $r_{\star}$ and $m_{\star}$ are stellar radius and mass, respectively. We introduce a new parameter called the penetration factor $\beta$, which is defined as $\beta \equiv r_\mathrm{t} / r_\mathrm{p}$, where $r_{\rm p}$ is the pericenter distance of the stellar orbit from the SMBH. Thus, a TDE occurs when $\beta >$ 1. 

After the destruction of a star, a spread in the specific energy of the stellar debris is produced by the tidal force of the SMBH \citep{rees88}:
\begin{equation}
    \label{eqn:spread}
    \Delta\epsilon\approx\frac{GM_{\rm SMBH}r_{\star}}{r_\mathrm{t}^2}.
\end{equation}

The possible range of the specific energy of the stellar debris is then
\begin{equation}
    \label{eqn:range}
    \epsilon_{\star}-\Delta\epsilon\leq\epsilon\leq\epsilon_{\star}+\Delta\epsilon,
\end{equation}
where $\epsilon_{\star}$ is the specific orbital energy of the star. 
\citet{Hayasaki_2018} uses this condition to classify TDEs into 5 types, based on the orbital eccentricities of tidally disrupted stars: (i) eccentric, (ii) marginally eccentric, (iii) parabolic, (iv) marginally hyperbolic, and (v) hyperbolic TDEs. Note that the term ``hyperbolic'' referring to the orbital nature of the incoming star should be distinguished from the classification of TDE types. Both marginally hyperbolic and hyperbolic TDEs arise from stars on hyperbolic orbits; however, the classification is determined by the fraction of stellar debris that remains bound to the SMBH after the disruption. The same distinction applies to the case of ``eccentric'' TDEs.

The parameter $\Delta\epsilon$ depends on the degree of disruption and the internal structure of the star. In a partial disruption, only the outer envelope may be stripped, leaving the dense stellar core intact. Consequently, the effective energy spread is expected to be smaller than the full-disruption estimate in \autoref{eqn:spread}. In this work, we therefore treat \autoref{eqn:spread} as an upper-limit estimate and leave a more detailed stellar-structure-dependent treatment for future work.

The bound mass of the stellar debris can be obtained by integrating the mass fallback rate. This rate is expressed as
\begin{equation}
    \frac{dm}{dt}=\frac{dm}{d\epsilon}\frac{d\epsilon}{dt},
\end{equation}
where $dm/d\epsilon$ is the mass distribution of the debris with respect to specific energy. Kepler’s third law gives
\begin{equation}
    d\epsilon/dt = (1/3)(2\pi G M_{\mathrm{SMBH}})^{2/3}t^{-5/3} \quad , 
\end{equation}
where $\epsilon=-GM_{\rm SMBH}/(2a)$ is the specific orbital energy of a bound debris element and $a$ is the semimajor axis of its orbit around the SMBH. Integrating the mass fallback rate yields
\begin{align}
    m_{\mathrm{bound}}
    &= \frac{m_{\star}}{2} \left[1 + \frac{\beta(1-e)}{2}\left(\frac{M_{\mathrm{SMBH}}}{m_{\star}}\right)^{1/3} \right], \label{eqn:bound}
\end{align}
where $e$ is the orbital eccentricity of the star (see \autoref{appendix:TDE_class} for details of the TDE classification and bound-mass calculation).

For neutron stars (NSs) and stellar-mass black holes (sBHs), we assume that the entire mass of each compact object is accreted by the SMBH once it reaches the accretion radius. For such compact objects, the tidal radius is not a physically appropriate criterion for capture, since relativistic effects become important \citep{Amaro-Seoanne2017LISA}. We therefore adopt this simplified treatment primarily to remain consistent with the current tidal disruption framework implemented in the code. Possible improvements to the treatment of compact object accretion are discussed in Section~\ref{sec:emri} and will be explored in future work.

Due to the relatively small particle number compared to real NSCs, a direct application of \hyperref[eqn:tidal]{Equation~\ref*{eqn:tidal}} is limited, as it would not produce any TDEs in such a system. Instead, we introduce an accretion radius $r_{\rm acc}$ in our \nbody{} code, which acts as a numerical criterion for tidal disruption. When a star reaches a distance $r \le r_{\rm acc}$ from the SMBH, it is assumed to be tidally disrupted and its debris is treated according to our accretion prescription. The parameter $r_{\rm acc}$ should therefore be regarded as a numerical capture radius rather than the true tidal radius given by \hyperref[eqn:tidal]{Equation~\ref*{eqn:tidal}}. Accordingly, the penetration factor $\beta$ is evaluated using $r_{\rm acc}$ in place of $r_{\rm t}$, such that $\beta = r_{\rm acc}/r_{\rm p}$. Following \citet{Kennedy2016}, we set the initial accretion radius to $r_{\rm acc}=3.00\times10^{-4}\,r_{\rm inf}$, where $r_{\rm inf}$ denotes the influence radius of the SMBH, defined as the radius within which the enclosed stellar mass equals the SMBH mass, i.e., 
\begin{equation}
    M_{\star}(<r_{\rm inf}) = M_{\rm SMBH}.
\end{equation}
This adopted $r_{\rm acc}$ is larger than the tidal radius of a Sun-like star in real galaxies by two to three orders of magnitude. Therefore, the absolute number of TDEs and the corresponding SMBH growth should not be interpreted as physical values without appropriate scaling, as the enlarged capture radius can overestimate the disruption rate. In this work, we use $r_{\rm acc}$ primarily to obtain sufficient statistics for comparing different accretion prescriptions under the same numerical setup, rather than to predict an absolute physical TDE rate.

For clarity, we distinguish three characteristic radii associated with the SMBH. The tidal radius $r_{\rm t}$ represents the true physical distance at which the tidal force of the SMBH disrupts a star. Since this distance scale is much smaller than the available spatial resolution in the present simulations, we instead employ the accretion radius $r_{\rm acc}$ as a numerical prescription for stellar removal and subsequent TDE processing within the \nbody\ framework. The influence radius $r_{\rm inf}$ characterizes the larger-scale dynamical region dominated by the SMBH and is used to normalize the adopted accretion radius.

In the FMA prescription adopted by \citet{panamarev2019}, $r_{\mathrm{acc}}$ remains fixed throughout the simulation, whereas our new PMA model allows it to expand as the SMBH grows through mass accretion from TDEs, following $r_{\rm acc} \propto M_{\rm SMBH}^{1/3}$. This scaling reflects the dependence of the physical tidal radius on SMBH mass. It therefore provides a simple way to couple the capture condition to the evolving SMBH instead of keeping it fixed to its initial value.

\subsection{Initial Conditions} \label{subsec:plummer}
We generate the initial particle distribution using a Plummer model \citep{plummer11, ahw74, Dejonghe_1987} consisting of $128\,000$ single stars.
We do not include primordial binaries in this first implementation, since the main goal of this pioneering study is to test the revised TDE accretion prescription. Stellar masses are drawn from the Kroupa initial mass function \citep{kroupa01} over the range from 0.08 to 150 \msun, yielding an average stellar mass of 0.59 \msun. The initial total stellar mass and half-mass radius of the NSC are $7.59\times10^{4}\,\msun$ and $0.78\,\mathrm{pc}$, respectively. We initialize the simulations with a zero-age stellar population and assume that the gas has been fully depleted following star formation. We adopt a metallicity of $Z=1.00\times10^{-3}$ and use the level-C prescriptions for stellar evolution, compact remnant formation, and natal kicks described by \citet{Kamlah2022}. In addition to the fixed central SMBH potential, the NSC is evolved within the external gravitational potential of a point-mass galaxy. The initial SMBH mass, $M_{\mathrm{SMBH,0}}$, is set to 10\% of the initial NSC mass, $M_{\rm tot}$. We further assume that the SMBH is non-rotating and does not host an accretion disc. After comparing these two accretion prescriptions, we perform two additional simulations using the PMA model with different initial SMBH masses. This allows us to examine how the depth of the central potential and the size of the loss-cone affect the TDE rate and the resulting SMBH growth. In total, we simulate three models with initial SMBH masses equal to 10\%, 15\%, and 20\% of the initial NSC mass.

To examine the effects of the improved TDE prescription without tying the results to a realistic physical NSC scaling, we adopt dimensionless \nbody\ (NB) units by setting the gravitational constant $G$ and the initial total stellar mass of the NSC, $M_{\mathrm{tot}}$, to unity, and the initial internal energy to $-0.25$ \citep{heggie_standardised_1986,Aarseth_Henon_Wielen_1974}. Here, $M_{\mathrm{tot}}$ excludes the SMBH, and the quoted initial energy refers to the isolated Plummer sphere before the fixed SMBH and external galactic potentials are included.

Each simulation is evolved for 2000 NB time units. We emphasize that the purpose of this work is not to reproduce a fully realistic NSC model, but to validate the improved TDE mass accretion treatment and its influence on SMBH growth within direct \nbody\ simulations. The chosen setup therefore provides a controlled framework for comparing different accretion prescriptions, despite the simplified SMBH treatment adopted in the simulations. Unless stated otherwise, all quantities in the following section are reported in NB units. A conversion table between NB units and physical units is provided in \autoref{appendix:unit}.

The simulations were performed on the \textsc{RAVEN} high-performance computing system at the Max Planck Computing and Data Facility (MPCDF) in Germany\footnote{\url{https://www.mpcdf.mpg.de/services/supercomputing/raven}}.

\section{Results} \label{sec:results}
\subsection{Comparison of Mass Accretion Treatments} \label{subsec:3.1}
In this section, we compare the FMA and PMA models to examine how the accretion treatment of disrupted stellar debris affects SMBH growth and NSC evolution.

\subsubsection{SMBH Growth via TDEs} \label{subsubsec:3.1.1}
\begin{figure}
    \centering
    \includegraphics[width=0.90\columnwidth]{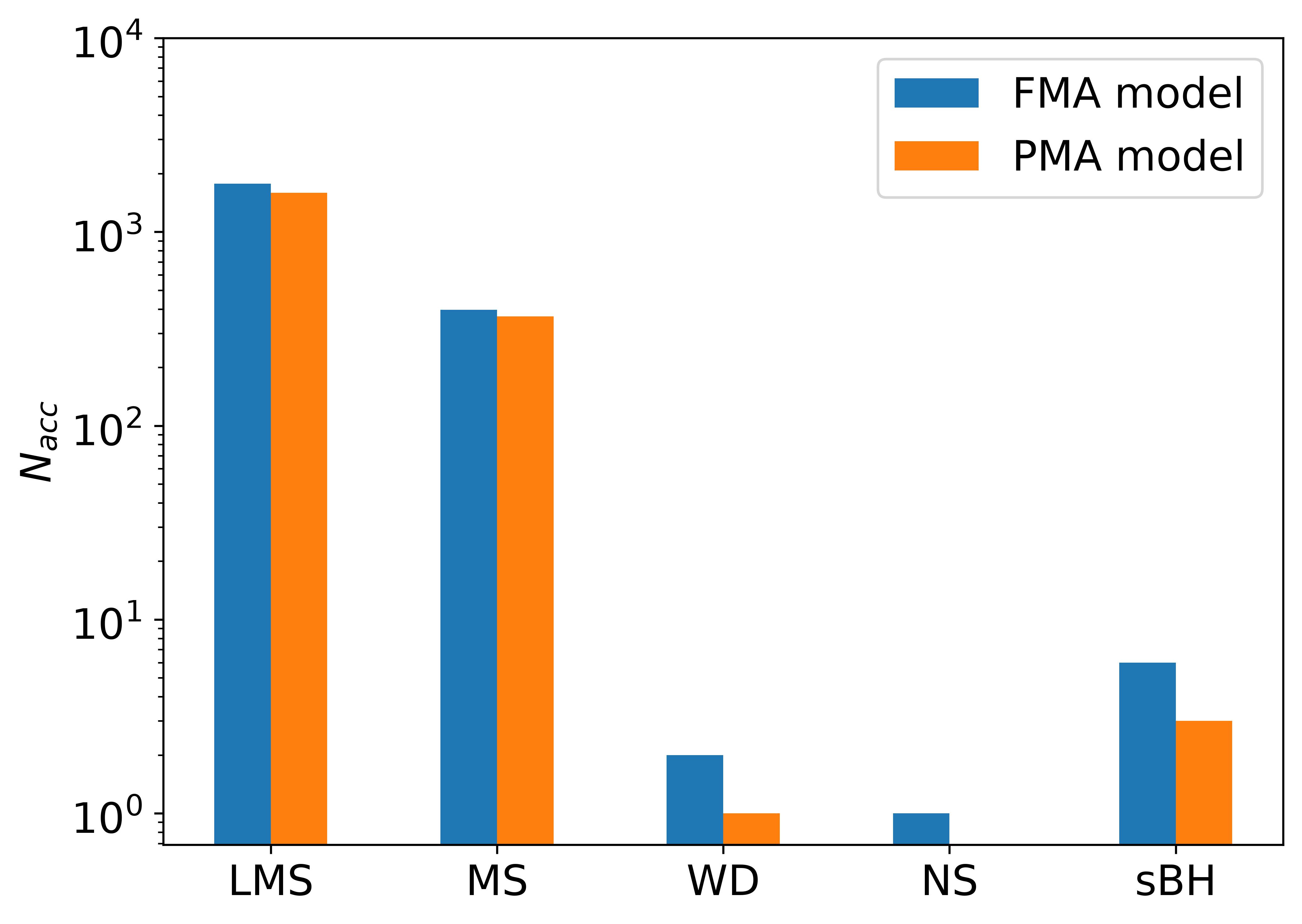}
    \caption{Number of accretion events, $N_{\rm acc}$, by stellar type in the FMA model (blue) and our PMA model (orange). The stellar types are low-mass main-sequence stars (LMS; $m_\star<0.7\,\msun$), main-sequence stars (MS; $m_\star\geq0.7\,\msun$), white dwarfs (WDs), neutron stars (NSs), and stellar-mass black holes (sBHs). Missing bars indicate no accretion events in the corresponding type. }
    \label{fig:type_dist1}
\end{figure}

To quantify how the two accretion prescriptions affect SMBH growth under otherwise identical initial conditions, we first compare the stellar types of the accreted objects in \autoref{fig:type_dist1}. The blue and orange bars represent the FMA and PMA models, respectively. Low-mass main-sequence (LMS) stars  account for more than 80\% of the TDEs in both models, while main-sequence (MS) stars constitute most of the remainder. Compact object captures are rare as only one white dwarf (WD) and three sBHs are captured in the PMA model, whereas the FMA model captures two WDs, one NS, and six sBHs. The dominance of MS stars reflects both the adopted initial mass function and the limited simulation duration, during which only a small fraction of the stellar population evolves beyond the MS.

\begin{figure*}[h]
    \centering
    \includegraphics[width=1.4\columnwidth]{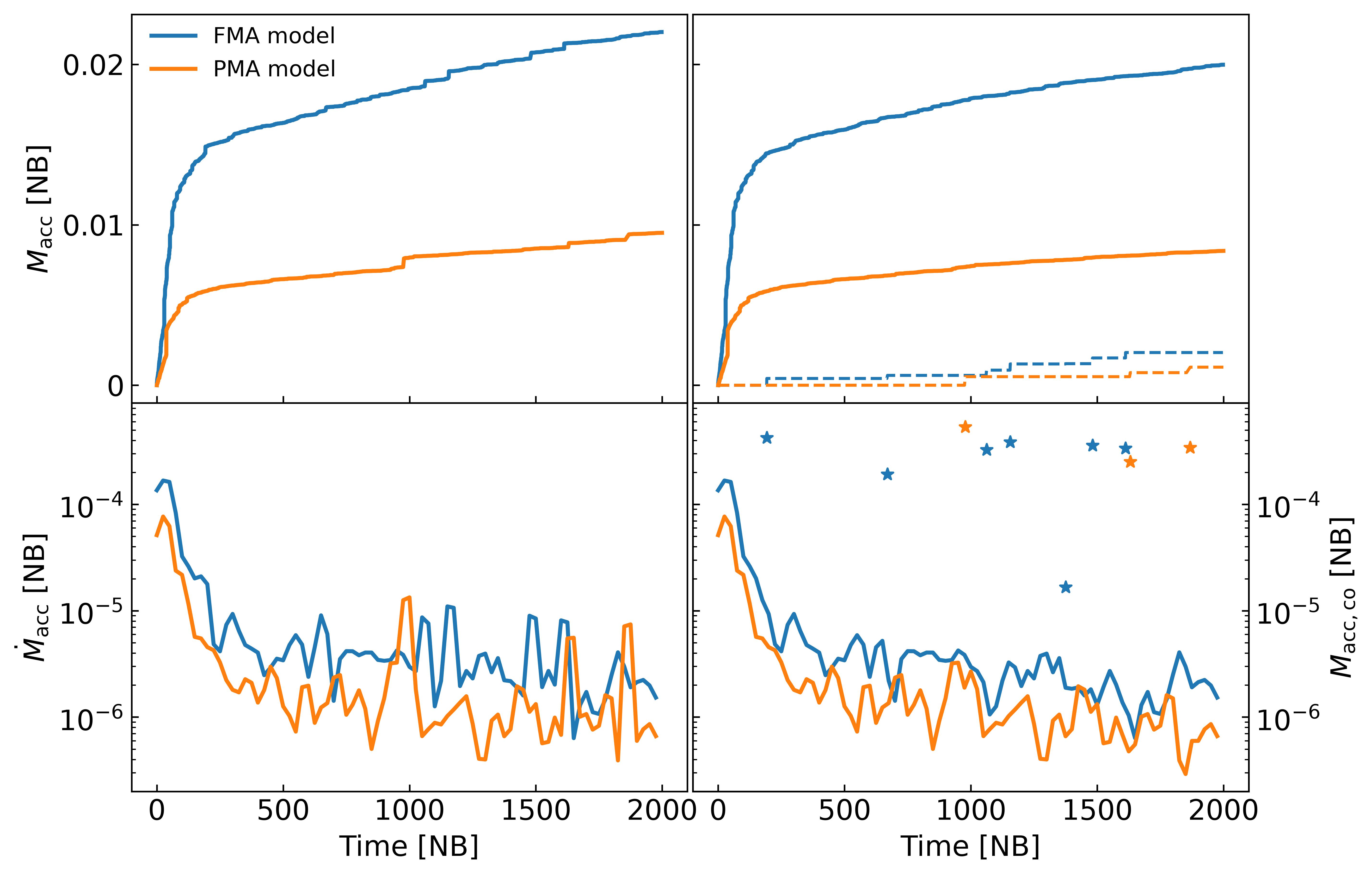}
    \caption{Total accreted mass (top) and mass accretion rate (bottom) as a function of time. The left panels include accretion from all particle types. In the right panels, the contribution from compact objects is separated from stellar TDEs. The cumulative mass accreted from compact objects is shown as dashed lines in the top-right panel, while individual compact object accretion events are indicated by star markers in the bottom-right panel, whose values correspond to the right-hand y-axis and indicate the mass of the accreted compact object.}
    \label{fig:m_acc_rate1}
\end{figure*}

\autoref{fig:m_acc_rate1} compares the cumulative accreted mass, $M_{\rm acc}$, and the mass accretion rate, $\dot{M}_{\rm acc}$, for the two prescriptions. Although the PMA model allows $r_{\rm acc}$ to grow as the SMBH gains mass, it produces fewer TDEs than the FMA model. This contrast suggests that the faster SMBH growth in the FMA model outweighs the effect of the growth of $r_{\rm acc}$ in the PMA model. In both models, $\dot{M}_{\rm acc}$ peaks at early times and then declines by more than an order of magnitude. This peak is associated with the rapid consumption of stars initially occupying the loss-cone and is further enhanced by the transient contraction during the cluster's adjustment to the SMBH potential (see \hyperref[subsubsec:3.1.2]{Section~\ref*{subsubsec:3.1.2}}). Once this initial population is depleted, two-body relaxation replenishes the loss-cone only slowly, so $\dot{M}_{\rm acc}$ remains low and declines only gradually.

The PMA model accumulates less mass throughout the simulation because it accretes only the bound fraction of the disrupted stellar mass. Individual compact object captures produce temporary spikes in $\dot{M}_{\rm acc}$. Their contribution to SMBH growth should be interpreted cautiously because only a few such events occur and the current prescription treats compact object captures as instantaneous without resolving their gradual inspiral toward the SMBH. When compact objects are excluded, $\dot{M}_{\rm acc}$ in the PMA model remains below that in the FMA model for nearly the entire evolution, confirming that the difference in SMBH growth is driven primarily by the treatment of stellar TDEs and the smaller bound mass assigned to each event in the PMA model.

\subsubsection{Evolution of the Cluster} \label{subsubsec:3.1.2}
\begin{figure}
    \centering
    \includegraphics[width=0.8\columnwidth]{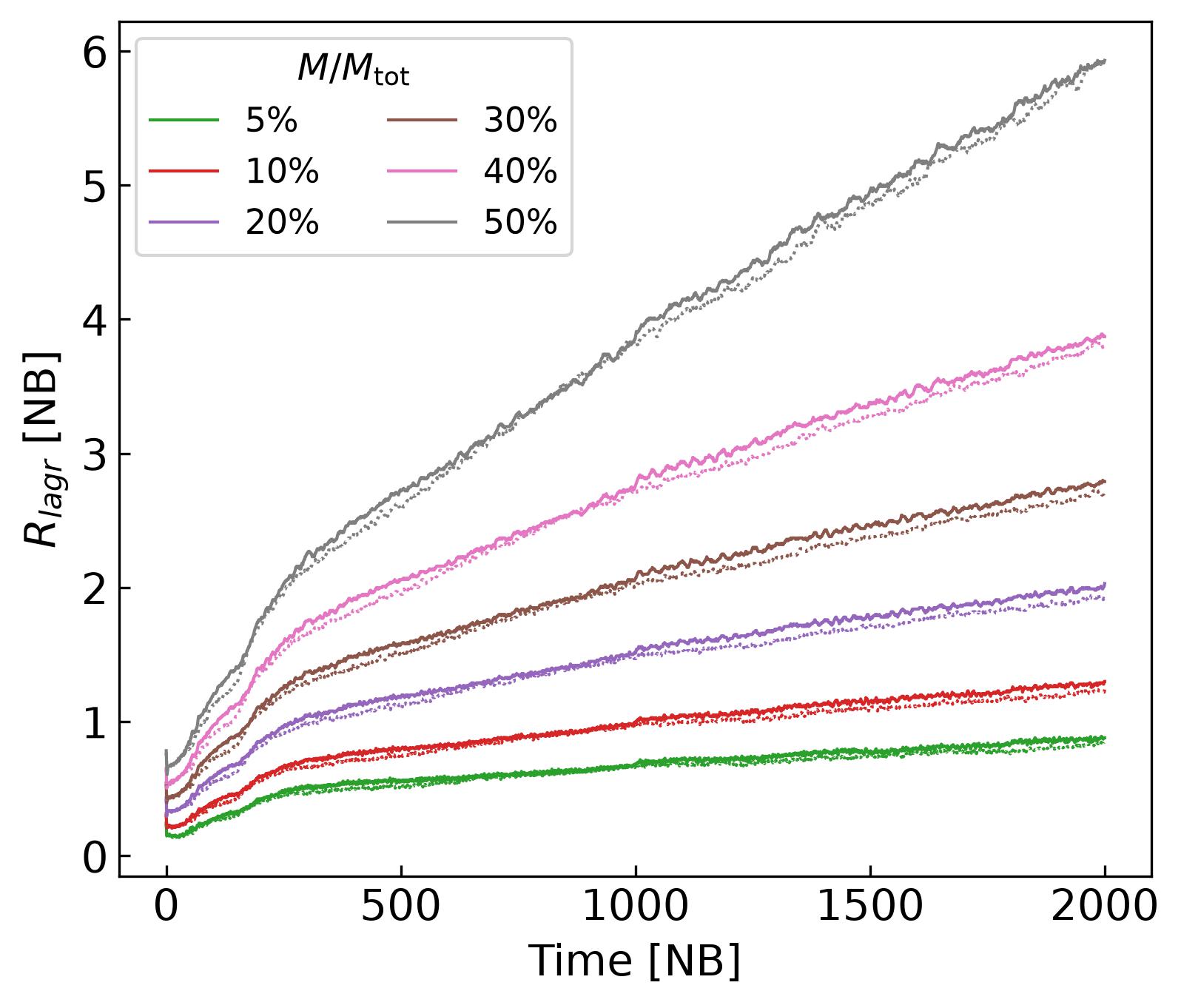}
    \caption{Evolution of the Lagrangian radii, calculated using the initial total stellar mass of the NSC. The dashed lines show the Lagrangian radii of the NSC from the FMA model, and the solid lines show those from the PMA model. The mass fraction corresponding to each Lagrangian radius is represented by a different color.}
    \label{fig:RLAGR}
\end{figure}

The dynamical evolution of the NSC is illustrated in \autoref{fig:RLAGR} by the Lagrangian radii, defined here as the radii enclosing fixed fractions of the initial total stellar mass. Initially, the NSC undergoes a rapid dynamical adjustment to the combined potential of the stellar system and the SMBH, causing it to contract. This initial adjustment occurs within approximately 1 NB time unit, corresponding to roughly one initial crossing timescale, as also indicated by the evolution of the virial ratio $Q$ in \autoref{fig:virial_ratio}.

After the initial adjustment, both clusters undergo sustained expansion, as indicated by the continuous increase in their Lagrangian radii throughout the simulation. This expansion is driven primarily by stellar mass loss due to stellar evolution. As shown in \autoref{fig:avmass}, the average stellar mass within both Lagrangian shells generally decreases after the early transient. The associated change in the cluster potential also perturbs the virial balance until $Q$ stabilizes near 0.5 at $t\sim1000$ NB, approximately halfway through the simulation. The initial increase in the average stellar mass within the inner (1--10\%) shell is mainly due to the early removal of stars through TDEs. Although the SMBH accretes considerably less mass under the PMA prescription, the Lagrangian radii follow similar evolutionary trends under the two prescriptions, with only minor differences in their expansion rates.

\begin{figure}
    \centering
    \includegraphics[width=0.95\columnwidth]{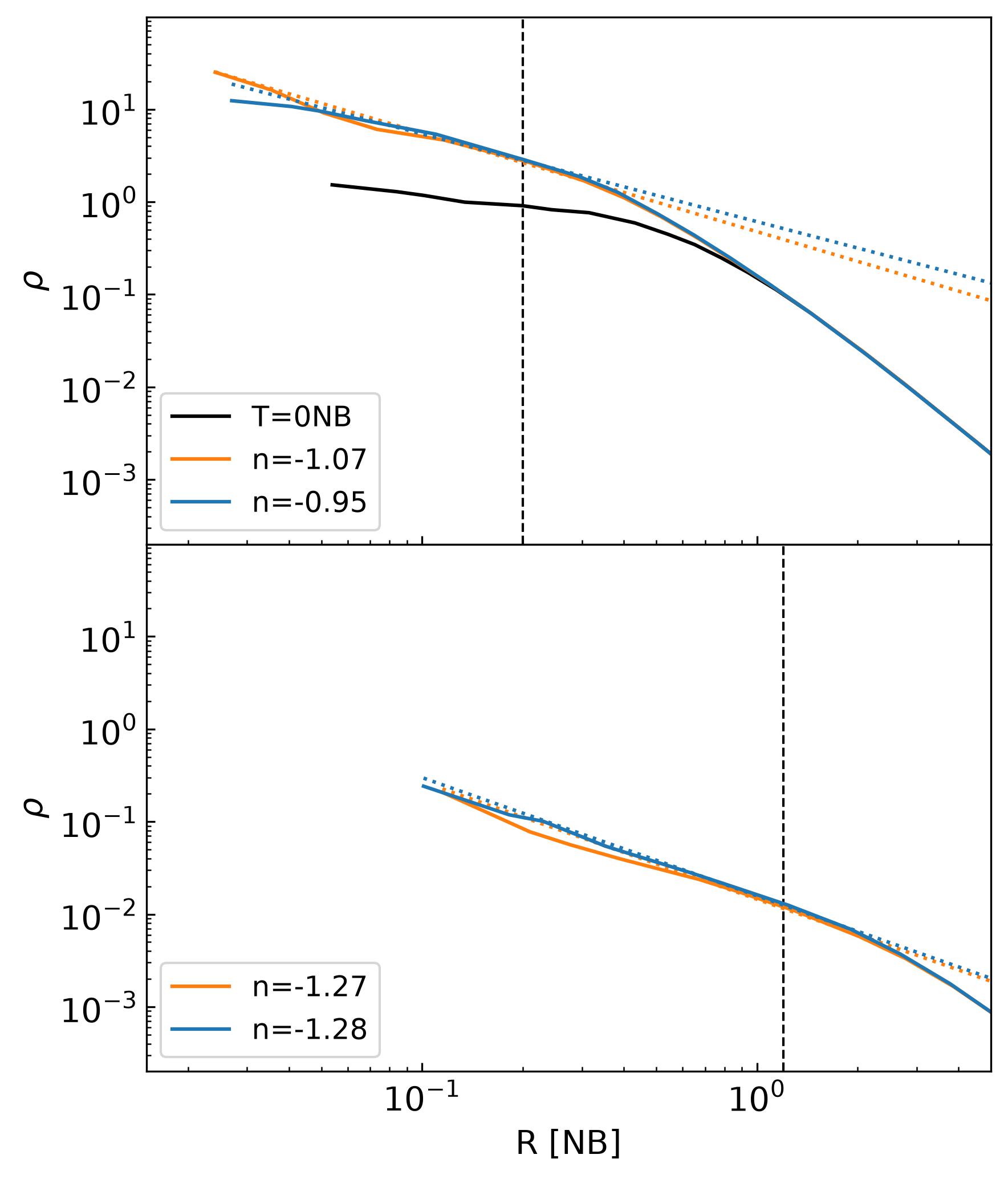}
    \caption{Density profiles of the FMA model (blue) and the PMA model (orange).
The initial profile at $T=0$ is shown as a solid black line in the top panel. The upper panel displays the profiles at $T=1\,\mathrm{NB}$, right after the initial adjustment, while the lower panel shows the profiles at the end of the simulation. Power-law fits within the influence radius are indicated by dotted lines. The influence radius at each epoch is marked by a vertical dashed line.}
    \label{fig:rho_profile}
\end{figure}

\autoref{fig:rho_profile} compares the stellar density profiles of the two models after the initial dynamical adjustment at $T=1\,\mathrm{NB}$ and at the end of the simulation. At $T=1\,\mathrm{NB}$, both simulations have developed central cusps within the influence radius, $R_{\rm inf}$, although their fitted slopes remain shallow. The PMA model has a slightly denser and steeper inner profile than the FMA model at this time, with fitted slopes of $-1.07$ and $-0.95$, respectively. Because the two models have accreted the same number of particles over this interval, this difference is consistent with the larger stellar mass removed per TDE under the FMA prescription. By the end of the simulation, both cusps have steepened and their fitted slopes are very similar, with values of $-1.27$ for the PMA model and $-1.28$ for the FMA model. These slopes remain shallower than the canonical Bahcall--Wolf value of $-1.75$ \citep{Bahcall_Wolf}. In a multi-mass system, different mass components can develop different cusp slopes, with the more massive components generally forming steeper distributions than low-mass stars \citep{panamarev2019}. The total stellar density profile, which is dominated by low-mass stars in our models, can therefore remain flatter.

\begin{figure}
    \centering
        \includegraphics[width=0.9\columnwidth]{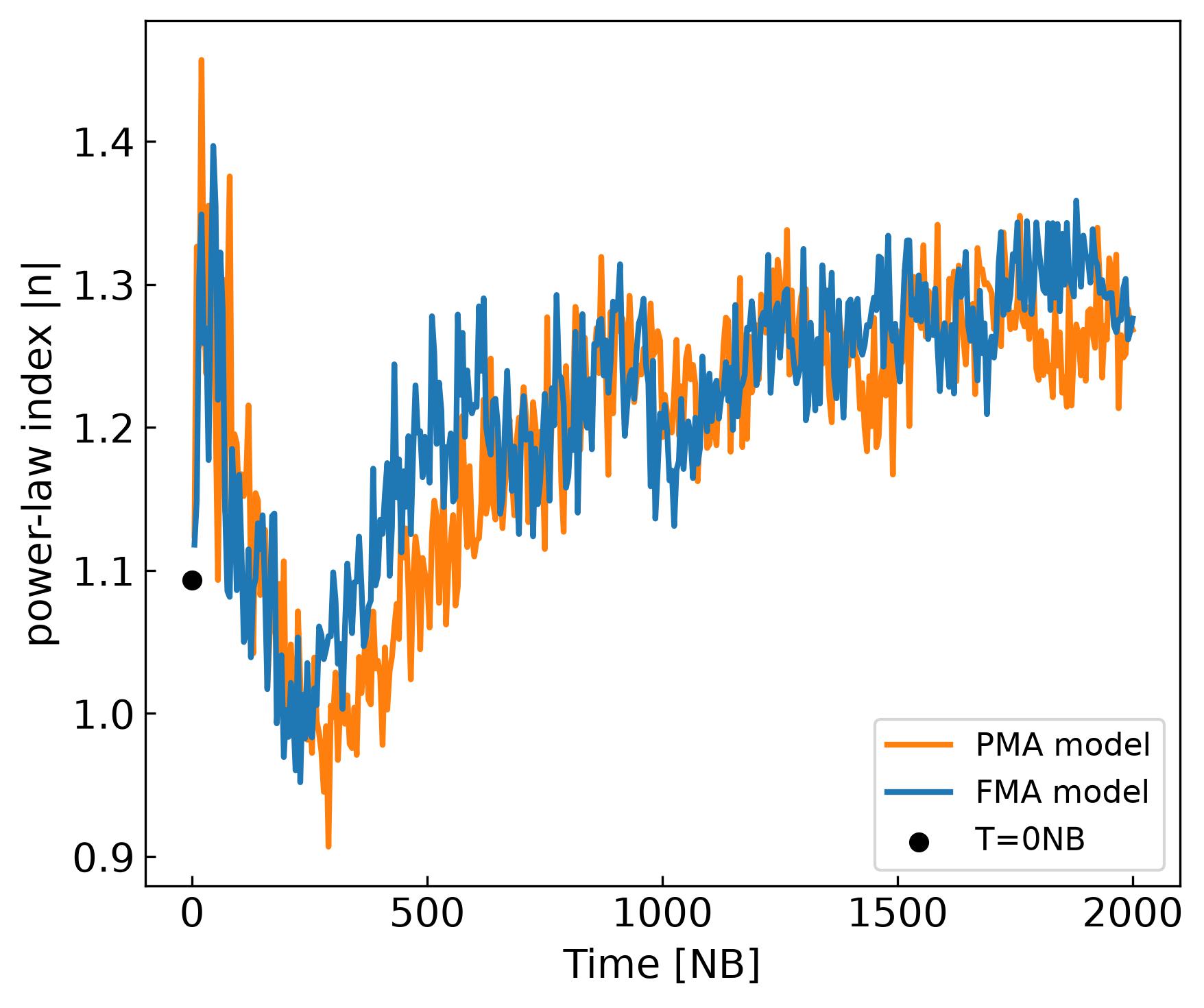}
    \caption{Time evolution of the absolute power-law index $|n|$ of the density profile inside the influence radius. The black point marks the initial value at the start of the simulation. The early spike and subsequent dip reflect the rapid dynamical adjustment of the cluster to the SMBH potential, while the later evolution shows the gradual steepening and fluctuation of the central cusp.
    }
    \label{fig:power-index}
\end{figure}

The time evolution of the inner density slope is shown in \autoref{fig:power-index}, where the absolute power-law index $|n|$ fitted within the influence radius is plotted as a function of time. The early spike and subsequent dip coincide with the rapid initial dynamical adjustment of the cluster to the SMBH potential. Thereafter, the cusp gradually steepens in both models, while $|n|$ exhibits moderate fluctuations. The two curves remain broadly similar and repeatedly overlap or cross, showing no persistent separation between the accretion prescriptions.

\subsection{Comparison of the PMA model with different ${M_{\rm SMBH,0}}$} \label{subsec:3.2}
In the preceding section, we assessed the effects of the improved accretion prescription by comparing the FMA and PMA models at a fixed initial SMBH mass. We now focus on the PMA model and compare three simulations with different initial SMBH masses. Because the initial SMBH mass determines the depth of the central potential and the size of the loss-cone, this comparison allows us to examine its influence on SMBH growth and TDE statistics.

\subsubsection{Growth of SMBHs} \label{subsubsec:growth}
\begin{figure*}
    \centering
    \includegraphics[width=1.4\columnwidth]{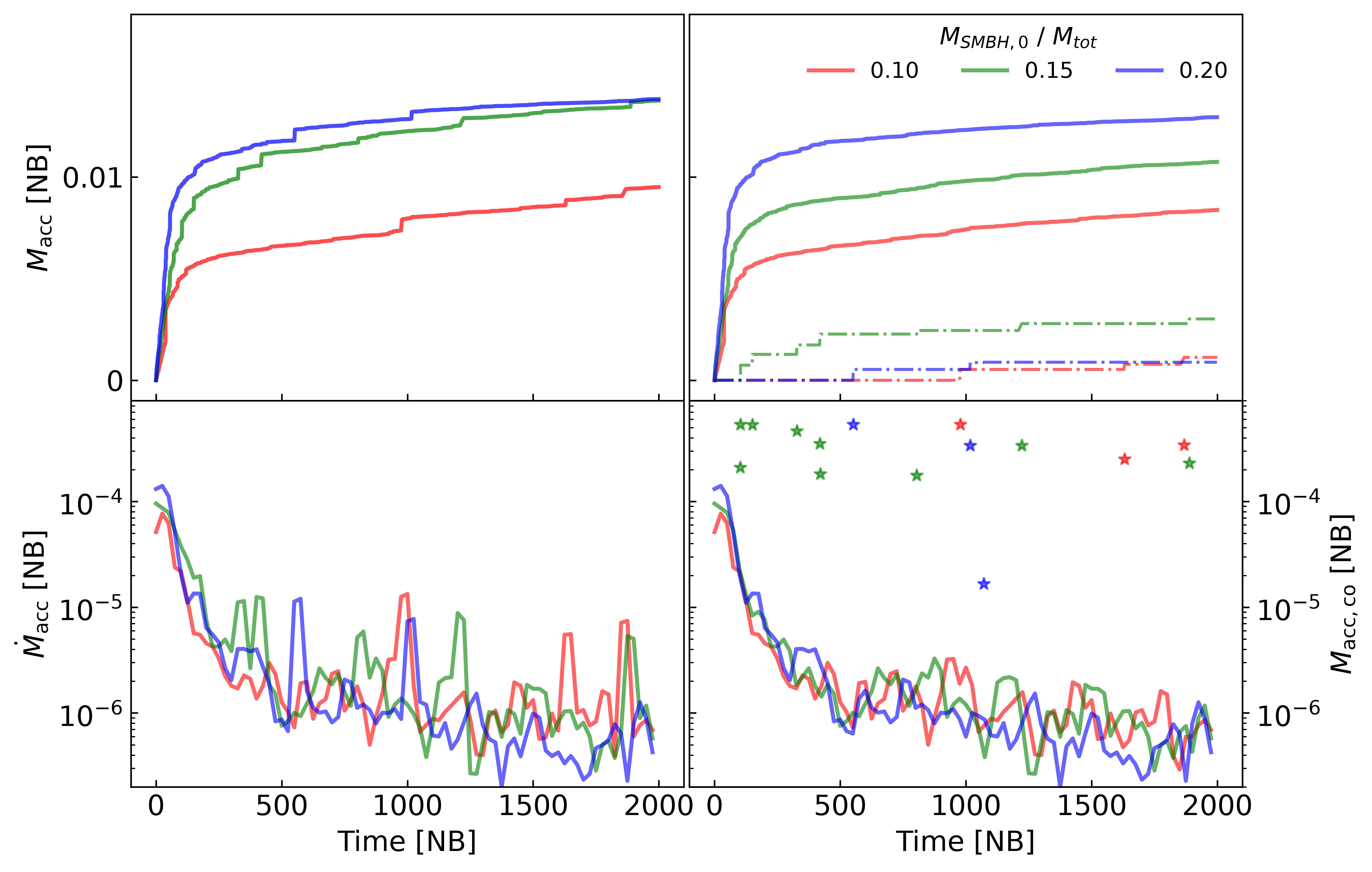}
    \caption{Total accreted mass (top panels) and mass accretion rate (bottom panels). The left panels show the accretion of all types. In the right panels, the solid lines represent the results after excluding the contribution from compact stars. The contribution from the compact objects is separated as dashed-dotted lines (top-right) and as 'star' markers (bottom-right).}
    \label{fig:m_acc_rate2}
\end{figure*}

We first examine how the initial SMBH mass affects SMBH growth by comparing the TDE-driven mass accretion histories shown in \autoref{fig:m_acc_rate2}. The left panels include accretion from all particle types, while the right panels separate the contribution from compact objects. In all three models, $\dot{M}_{\rm acc}$ peaks at early times and subsequently declines by more than an order of magnitude. This early peak reflects the rapid consumption of stars from the full loss-cone. After this population is depleted, further accretion is limited by the slower replenishment of the loss-cone through two-body relaxation.

The cumulative accreted mass generally increases with the initial SMBH mass, with the $M_{\rm SMBH,0}/M_{\rm tot}=0.10$ model accreting the least mass and the $M_{\rm SMBH,0}/M_{\rm tot}=0.20$ model accreting the most. The total accreted masses of the $M_{\rm SMBH,0}/M_{\rm tot}=0.15$ and $0.20$ models are similar, but their ordering becomes clearer when compact objects are excluded, indicating that the overall trend is driven primarily by TDEs. Individual compact object captures produce discrete increases in the cumulative accreted mass and partially obscure this trend. However, because these events are rare, sensitive to stochastic sampling of the initial conditions, and treated with a simplified accretion prescription that neglects gradual inspiral, their differences among the models should not be interpreted as a systematic dependence on the initial SMBH mass. 

The higher early-time mass accretion rates in models with more massive initial SMBHs are consistent with their larger $r_{\rm acc}$ and more extensively populated loss-cones. Although $r_{\rm acc}$ continues to grow as the SMBH accretes mass, this growth does not sustain the initially enhanced $\dot{M}_{\rm acc}$. Instead, the rates in the higher-mass models decline more rapidly, reducing the differences among the three models. The rates overlap quickly after the peak and thereafter fluctuate around $\dot{M}_{\rm acc}\sim10^{-6}$ in NB units, with no persistent ordering among the models. This convergence indicates that the growth of $r_{\rm acc}$ alone cannot maintain a higher accretion rate after the initially available loss-cone population has been depleted.

\begin{figure}
    \centering
    \includegraphics[width=1.0\columnwidth]{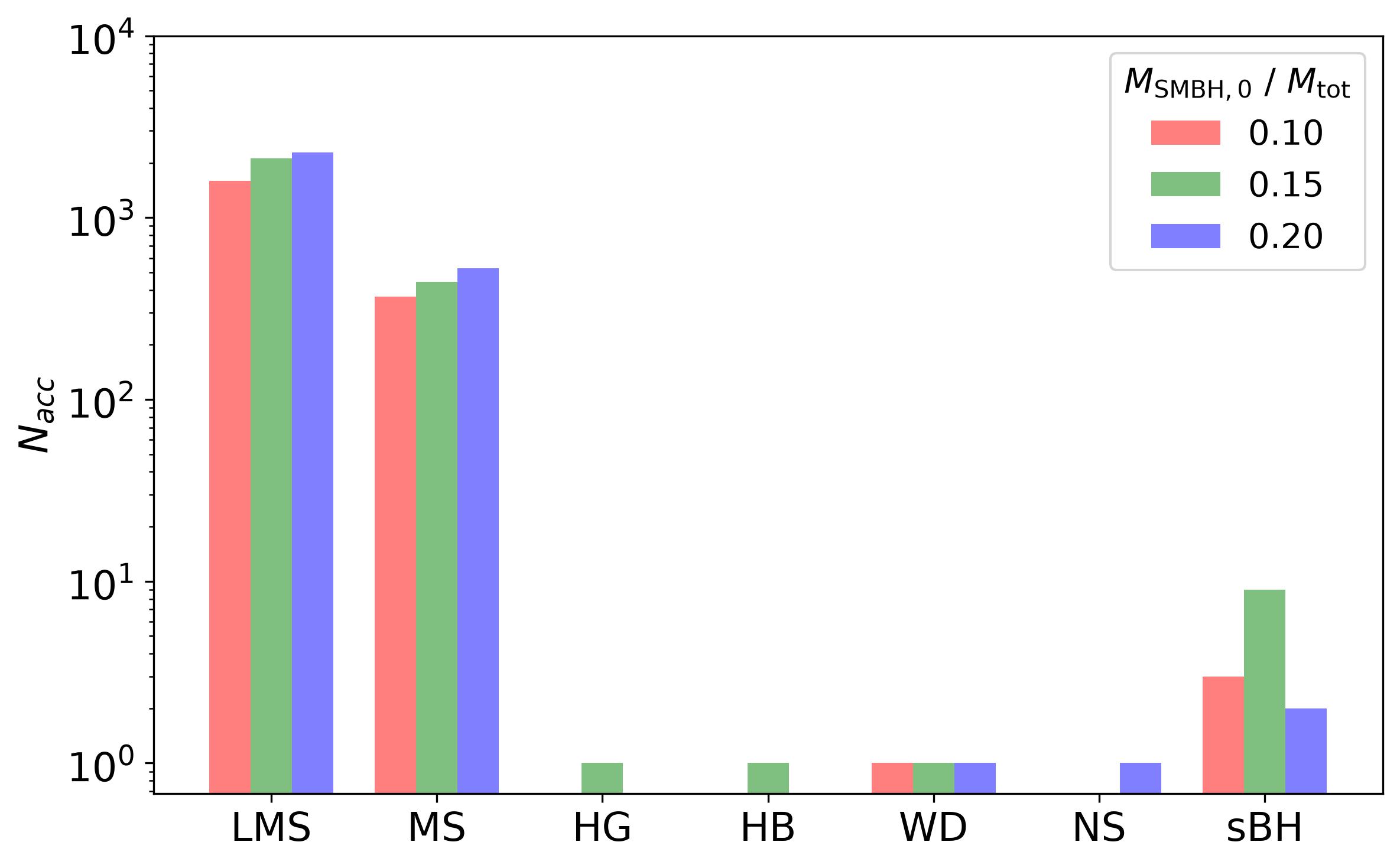}
    \caption{Number of accretion events, $N_{\rm acc}$, by stellar type in the PMA models with $M_{\rm SMBH,0}/M_{\rm tot}=0.10$ (red), $0.15$ (green), and $0.20$ (blue). HG and HB denote Hertzsprung-gap and core-helium-burning stars, respectively; the other stellar-type abbreviations are defined in \autoref{fig:type_dist1}.}
    \label{fig:type_dist2}
\end{figure}

The stellar type distribution of accreted particles is shown in \autoref{fig:type_dist2}. LMS stars dominate the accreted population in all three models, followed by MS stars, broadly reflecting the adopted Kroupa initial mass function. The numbers of accreted LMS and MS stars both increase with $M_{\rm SMBH,0}$. By contrast, evolved-star and compact object captures are rare and do not exhibit a monotonic dependence on the initial SMBH mass. Each model captures one WD, while the $M_{\rm SMBH,0}/M_{\rm tot}=0.10$, $0.15$, and $0.20$ models capture three, nine, and two sBHs, respectively. Only the $M_{\rm SMBH,0}=0.15M_{\rm tot}$ model captures a Hertzsprung-gap star and a core-helium-burning star, and only the $M_{\rm SMBH,0}=0.20M_{\rm tot}$ model captures an NS. Given these small numbers, the differences among the models should not be interpreted as systematic trends. The accreted population is therefore overwhelmingly composed of unevolved stars.  Since the main-sequence lifetimes of LMS can exceed $\sim 1\,\mathrm{Gyr}$, substantially longer integrations are required to assess whether post-main-sequence stars can contribute significantly to the TDE statistics and to SMBH growth. In addition, disruptions of evolved stars require a prescription that accounts for their centrally concentrated structure, because their envelopes may be stripped while their cores survive.

\subsubsection{TDE Statistics} \label{subsubsec:dist}

\begin{table}[h]
\centering
{\footnotesize
\begin{tabular}{c||ccccccccc} 
 \hline \\[-8pt]
    $M_{\mathrm{SMBH,0}} / M_{\mathrm{tot,0}}$ & $N_{\mathrm{acc}}$ &
    $f_{\mathrm{e}}$  & $f_{\mathrm{me}}$ & $f_{\mathrm{p}}$ &
    $f_{\mathrm{mh}}$ & $f_{\mathrm{h}}$ & $N_{\mathrm{co}}$ \\ [1pt]
 \hline \\[-8pt]
    0.10 & 1963 & 0.3 & 68.2 & 0 & 31.5 & 0 & 3 \\
    0.15 & 2576 & 0.8 & 66.0 & 0 & 33.2 & 0 & 9 \\
    0.20 & 2819 & 0.5 & 68.3 & 0 & 31.2 & 0 & 3 \\
 \hline 
\end{tabular}
} \vspace{5pt}
\caption{Statistics of TDE types. $N_{\mathrm{acc}}$ denotes the total number of accretion events, including NS and sBH captures. The quantities $f_{\mathrm{e}}$, $f_{\mathrm{me}}$, $f_{\mathrm{p}}$, $f_{\mathrm{mh}}$, and $f_{\mathrm{h}}$ denote the percentages of stellar TDEs classified as eccentric, marginally eccentric, parabolic, marginally hyperbolic, and hyperbolic, respectively, calculated after excluding NS and sBH captures. $N_{\mathrm{co}}$ denotes the number of NS and sBH captures.}
\label{tab:result}
\end{table}

We now compare the statistical properties of TDEs among the three simulations to assess their dependence on the initial SMBH mass. \autoref{tab:result} summarizes the number of accretion events and the relative frequencies of the TDE types. Columns 3--7 give the percentages of eccentric, marginally eccentric, parabolic, marginally hyperbolic, and hyperbolic TDEs, respectively. NS and sBH captures are excluded from these fractions because they are treated using the simplified prescription described in \hyperref[subsubsec:growth]{Section~\ref*{subsubsec:growth}}. Marginally eccentric and marginally hyperbolic TDEs dominate in all three models and together account for more than $99\%$ of the stellar disruptions, consistent with previous numerical results \citep{Hayasaki_2018}. Eccentric TDEs constitute only $0.3$--$0.8\%$ of the events, and no parabolic or hyperbolic TDEs are identified. The absence of strictly parabolic events is expected because this class requires the exact condition $e=1$, which is rarely realized in numerical simulations and is also unlikely in realistic stellar systems.

\begin{figure*}
    \centering
    \includegraphics[width=1.0\textwidth]{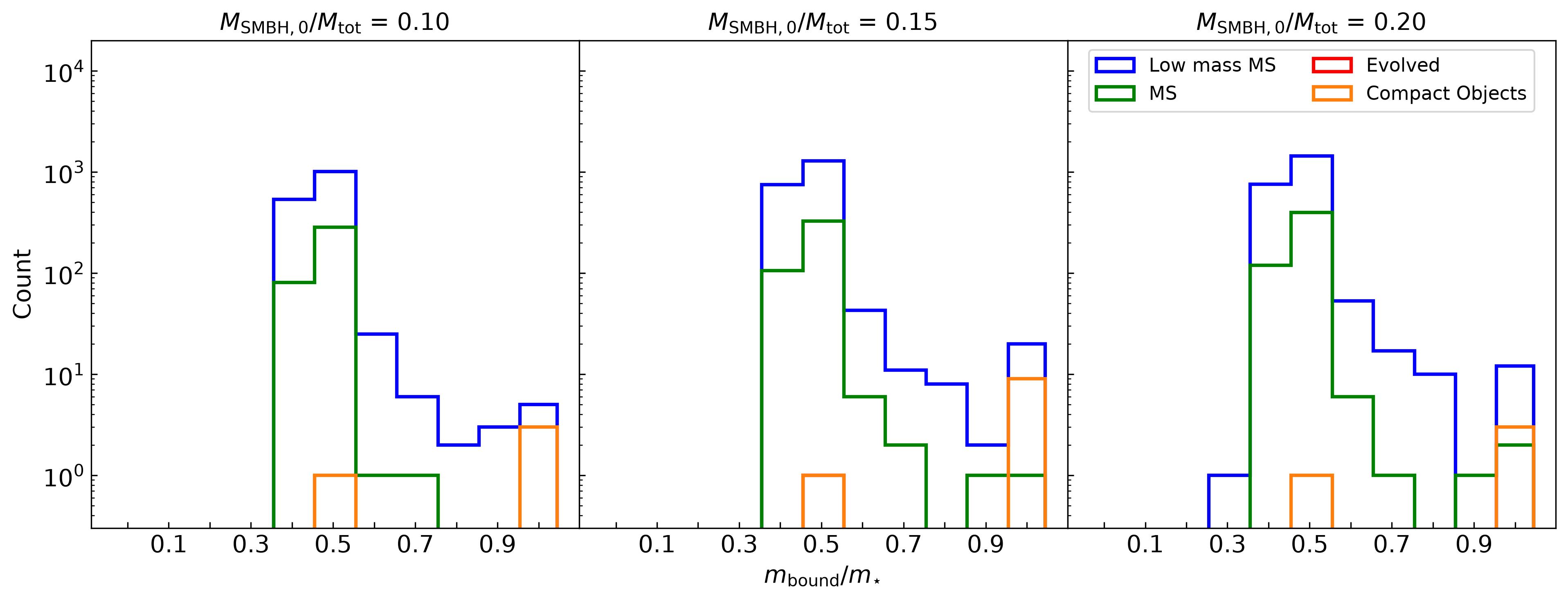}
    \caption{Histogram of the bound debris mass fraction, $m_{\rm bound}/m_{\star}$, for all TDEs. Panels show results for different initial SMBH masses, while colors indicate different stellar types (LMS, MS, evolved stars, and compact objects).} 
    \label{fig:M_frac_dist}
\end{figure*}

\autoref{fig:M_frac_dist} shows the distribution of the bound debris mass fraction, $m_{\rm bound}/m_{\star}$, for stellar TDEs, together with compact object captures assigned a fully accreted mass fraction. In all three models, at least $\sim95\%$ of the stellar TDEs lie within $0.4\lesssim m_{\rm bound}/m_{\star}\lesssim0.6$, consistent with the predominance of nearly parabolic encounters. The distributions also contain a sparsely populated tail toward larger bound mass fractions. In particular, marginally eccentric TDEs increasingly include events with higher $m_{\rm bound}/m_{\star}$ for larger $M_{\mathrm{SMBH,0}}$, but the eccentric-TDE fraction does not increase monotonically with the initial SMBH mass. According to \autoref{eqn:ecrit1}, increasing the SMBH-to-star mass ratio moves $e_{\rm crit,1}$ closer to unity and can therefore enlarge the range of bound stellar orbits classified as eccentric TDEs. However, the measured fraction also depends on the sampled orbital distribution and remains too small to establish a monotonic trend.

\begin{figure*}
    \centering
        \includegraphics[width=1.0\textwidth]{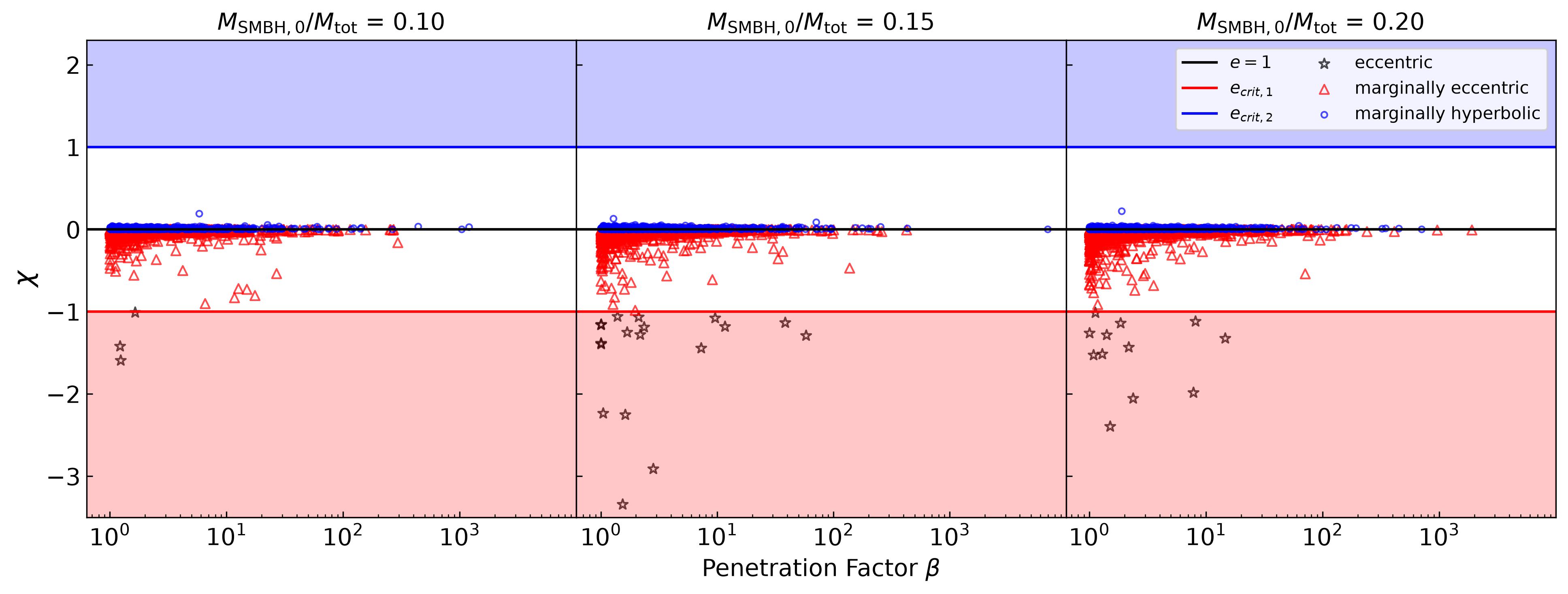}
    \caption{TDE distribution of the dimensionless parameter $\chi$ as a function of the penetration factor $\beta$ for TDEs with different initial SMBH masses. Horizontal lines at $\chi = -1$, 0, and 1 indicate the boundaries between eccentric, parabolic, and hyperbolic TDE regimes. Symbols denote different TDE types.}
    \label{fig:chi-beta}
\end{figure*}

\autoref{fig:chi-beta} shows the distribution of TDE types in the $\chi$--$\beta$ plane. The dimensionless parameter $\chi$ combines the orbital eccentricity, penetration factor, and SMBH-to-star mass ratio, allowing the types to be separated by the fixed boundaries $\chi=-1$, 0, and 1. Its definition and the classification criteria are given in \autoref{appendix:TDE_class}. Most events cluster close to $\chi=0$ on either side of the parabolic boundary. Accordingly, more than $99\%$ of the stellar TDEs are marginally eccentric or marginally hyperbolic, for which the bound debris fraction is approximately one-half. The relative frequencies of these two dominant types vary only slightly among the models. Eccentric TDEs with $\chi<-1$ remain rare, and the negative-$\chi$ tail does not broaden monotonically with the initial SMBH mass; among the three simulations, the $M_{\rm SMBH,0}/M_{\rm tot}=0.15$ model extends to the most negative values of $\chi$. The initial SMBH mass therefore has a clearer effect on the total number of TDEs than on their relative distribution among orbital types.

\section{Improving the Treatment of Compact Object Accretion} \label{sec:emri}
In addition to stellar tidal disruption events, NSCs also contain stellar-mass compact objects which require a different treatment when they approach the SMBH. Unlike normal stars, compact objects are not tidally disrupted at the tidal radius. In many cases their tidal radius lies within the event horizon of the SMBH, implying that they cannot be tidally torn apart before crossing the horizon \citep{amaro04}. As a result, the mechanism governing their capture differs fundamentally from stellar TDEs. Instead, compact objects typically undergo a gradual orbital decay toward the SMBH. During this process, the compact object repeatedly passes through periapsis and slowly loses orbital energy through the emission of gravitational waves (GWs). This phenomenon is known as an extreme-mass ratio inspiral (EMRI), reflecting the large mass ratio between the compact object and the SMBH, typically $m_{\rm CO}/M_{\rm SMBH} \sim 10^{-8} - 10^{-5}$ \citep{AmaroSeoane_2022}. As the orbit shrinks, the compact object can undergo up to tens of thousands to millions of orbital cycles before eventually falling into the SMBH.

Because EMRIs evolve over many orbital periods rather than through a single disruption event, modeling compact object accretion using a simple accretion radius in our \nbo{} code is not appropriate. Such an approach would lead to artificially frequent accretion events and an overestimate of the contribution of compact objects to SMBH growth. More realistic treatments therefore attempt to incorporate relativistic effects and the gradual orbital decay driven by GW emission. 

To approximate this relativistic inspiral, \citet{Minzburg2022} introduced a capture criterion based on the gravitational radiation timescale derived from the orbit-averaged formalism of \citet{PetersMathews1963} and \citet{Peters1964}. This represents one of the first attempts to incorporate EMRIs into the direct \nbo{} code, although the treatment remains highly simplified. In this approach, the inspiral of a compact object of mass $m$ around an SMBH of mass $M$ is driven by the GW emission. The orbit-averaged decay of the semi-major axis is given by
\begin{equation}
\left\langle \frac{da}{dt} \right\rangle = -\frac{64}{5} \frac{G^3 m M (m+M)}{c^5 a^3 (1-e^2)^{7/2}}
\left(1+\frac{73}{24}e^2+\frac{37}{96}e^4\right),
\end{equation}
where $a$ and $e$ denote the semi-major axis and eccentricity of the orbit, respectively. The corresponding GW inspiral timescale is then approximated as
\begin{equation}
t_{\rm gr,bound} \equiv \frac{a}{|\dot{a}|} = \frac{5}{64}\frac{c^5 a^4}{G^3 m M (m+M) f(e)},
\end{equation}
with
\begin{equation}
f(e)=\frac{1+\frac{73}{24}e^2+\frac{37}{96}e^4}{(1-e^2)^{7/2}}.
\end{equation}

Since the Peters (1964) inspiral timescale is only valid for bound orbits ($e<1$), a separate treatment is required for hyperbolic orbits. \citet{Minzburg2022} derived an analogous GW inspiral timescale for hyperbolic encounters,
\begin{equation}
    t_{\rm gr,unbound} = \frac{5}{64}
    \frac{c^5 |a|^4 (e^2-1)^{7/2}}
    {G^3 mM(m+M)\tilde{G}(e)},
\end{equation}
where $\tilde{G}(e)$ is given by the following expression \citep{Turner_1977,DeVittori_2012}:
\begin{equation*}
    \tilde{G}(e) = \frac{1}{\pi} \cos^{-1}\left(-\frac{1}{e}\right)f(e)(1-e^2)^{7/2}
    +\frac{(e^2-1)^{1/2}}{24\pi}\left(\frac{301}{6}+\frac{673}{12}e^2\right)
\end{equation*}

To determine whether GW emission dominates the orbital evolution, the inspiral timescale is compared to the Keplerian orbital period,
\begin{equation}
t_{\rm orb} = 2\pi \sqrt{\frac{a^3}{G(m+M)}}.
\end{equation}
Whenever a compact object enters the influence radius of the SMBH, $t_{\rm gr}$ and $t_{\rm orb}$ are evaluated from its instantaneous orbital elements. If $t_{\rm gr} < t_{\rm orb}$, the orbit is expected to decay within one orbital timescale due to GW emission, and the compact object is assumed to be accreted onto the SMBH. This prescription provides a simple approximation for the onset of an EMRI without explicitly integrating the relativistic orbital evolution.

To test the compact object accretion prescription, \citet{Minzburg2022} performed a pilot simulation consisting exclusively of $100\,000$ sBH particles. In that simulation, the cluster initially contained only compact objects, allowing compact object capture through GW-driven inspiral to be studied separately from stellar TDEs. The simulation followed the dynamical evolution of the sBHs in the SMBH potential and applied the capture criterion described above to determine when they merged with the SMBH.

\begin{figure}
    \centering
    \includegraphics[width=1.0\columnwidth]{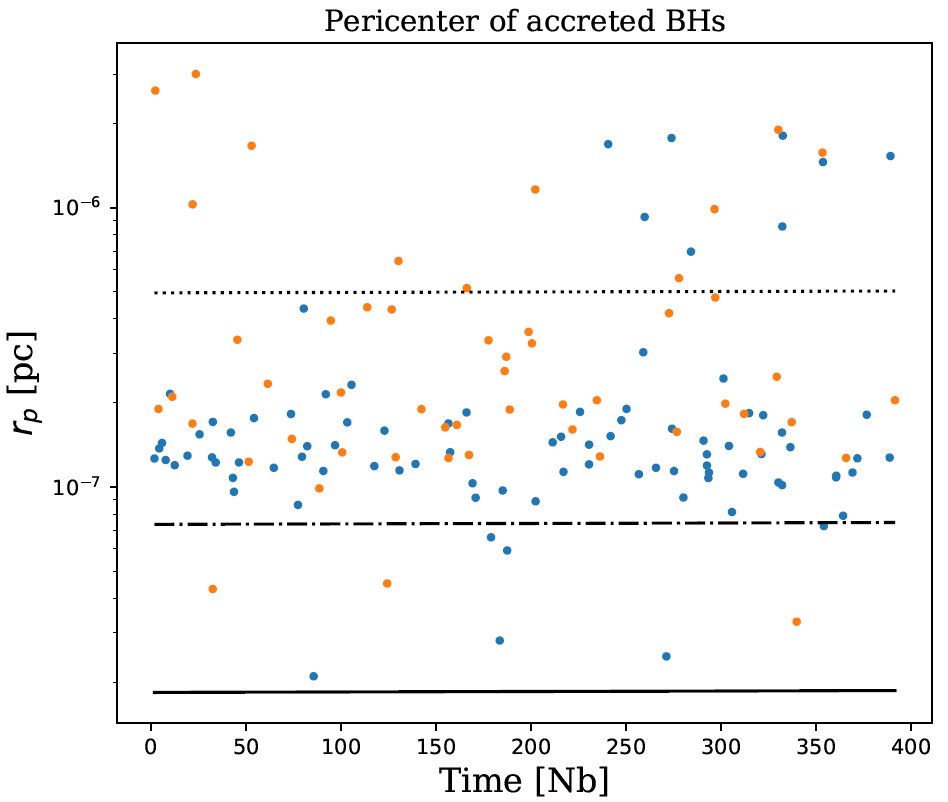}
    \caption{Pericenter of accreted black holes. Blue and orange dots represent those with eccentric and hyperbolic orbits, respectively. The solid line represents the Schwarzschild radius, $r_{\rm s}$,  of the SMBH, the dashed-dotted line is $4r_{\rm s}$, and the dotted line is $27r_{\rm s}$. Reprinted from \citet{Minzburg2022}.}
    \label{fig:pericenter_bh}
\end{figure}

The results of this pilot simulation show that most accreted black holes approach the SMBH on highly eccentric orbits with small pericenter distances. Figure~\ref{fig:pericenter_bh} illustrates the pericenter distance of accreted black holes as a function of time. The majority of accretion events occur for pericenter distances between approximately $4$ and $27$ Schwarzschild radii of the SMBH, indicating that the compact objects reach very deep regions of the potential before merging. Only a small number of outliers exhibit larger pericenter distances, which are typically associated with nearly parabolic orbits. These results demonstrate that the new EMRI criterion may preferentially select compact objects that undergo deep plunges toward the SMBH.


\section{Conclusions} \label{conclusion}
In this work, we developed and tested an improved prescription for the accretion of stellar debris following TDEs in the direct $N$-body code \textsc{Nbody6++GPU}. Using simulations of NSCs hosting a central SMBH, we investigated how the adopted accretion prescription affects SMBH growth and NSC evolution. Our main conclusions are as follows:

\begin{itemize}
    \item In the PMA model, only the gravitationally bound fraction of the stellar debris is added to the SMBH, rather than the entire mass of the disrupted star. The debris energy distribution is used to classify TDEs and determine their bound mass fractions, while compact object captures retain the simplified full-accretion prescription.

    \item Compared with the FMA model, the PMA model produces fewer TDEs, lower mass accretion rates, and less cumulative SMBH growth, even though $r_{\rm acc}$ in the PMA model is allowed to increase as the SMBH grows. By assigning the entire disrupted stellar mass to the SMBH, the FMA prescription overestimates the mass accreted per TDE relative to the PMA prescription. The greater number of TDEs in the FMA model further increases the difference in cumulative SMBH growth.

    \item Both models undergo a rapid initial contraction followed by sustained expansion. The subsequent expansion is driven primarily by stellar mass loss. Although the SMBH accretes considerably less mass in the PMA model, the Lagrangian radii evolve similarly under the two prescriptions, with only minor differences in their expansion rates.

    \item For initial SMBH masses equal to 10\%, 15\%, and 20\% of the initial cluster mass, the total number of stellar TDEs and the cumulative accreted mass generally increase with $M_{\rm SMBH,0}$. Models with higher initial mass accretion rates show a more rapid subsequent decline, and the rates overlap soon after the peak, with no persistent ordering at later times.

    \item Regardless of the initial SMBH mass, more than 99\% of stellar TDEs are marginally eccentric or marginally hyperbolic, indicating that most disrupted stars approach the SMBH on nearly parabolic orbits. Consequently, at least approximately 95\% of the events have bound debris fractions in the range $0.4 \lesssim m_{\rm bound}/m_{\star} \lesssim 0.6$. The initial SMBH mass therefore has a clearer effect on the total number of TDEs than on the distribution of TDE types.

    \item LMS and MS stars dominate the accreted population in all simulations, whereas evolved-star disruptions and compact object captures are rare under the adopted initial conditions and simulation duration. The central stellar density cusps steepen with time but remain shallower than the canonical Bahcall--Wolf slope. This is consistent with the total stellar density profiles being dominated by low-mass stars, whereas more massive components are expected to develop steeper distributions.
\end{itemize}

Several limitations of the present modeling should be considered when interpreting these results. We represent the unresolved physical tidal radius using an enlarged numerical accretion radius, $r_{\rm acc}$, to obtain sufficient disruption events for a controlled comparison of the two prescriptions. Future implementations should allow this numerical criterion to depend on stellar mass, radius, and internal structure. This refinement is particularly important for partial disruptions of evolved stars, in which the envelope may be stripped while the core survives with a modified orbital energy \citep{Ryu_2020,Manukian+2013,Chen+2024ApJ}. Compact object captures also require a more realistic treatment because the current prescription assumes instantaneous full accretion and does not resolve gradual relativistic inspiral.

Our simulations also represent a single cluster mass scale with substantially fewer particles than real NSCs. Together with the enlarged numerical accretion radius, this limitation means that the absolute TDE counts and corresponding SMBH growth should not be interpreted as direct predictions for observed NSCs \citep{Zhong_2014, Zhong_2015, panamarev2019, Li_2019}. We therefore restrict our interpretation to comparisons among models evolved under the same numerical conditions. Future simulations with larger particle numbers will allow more realistic studies of NSC dynamics, TDEs, and EMRIs over longer timescales. The ongoing DRAGON-III project, for example, aims to model NSCs with up to one million particles using \nbo{} \citep{Wu.et.al_2025}.

\begin{acknowledgements}
PC is a fellow of the International Max Planck Research School for Astronomy and Cosmic Physics at Heidelberg University (IMPRS-HD) and acknowledges support from DAAD (funding program number 57693453). PC also acknowledges computing time granted by the John von Neumann Institute for Computing (NIC) on the GCS Supercomputer JUWELS Booster, as well as support from the Max Planck Computing and Data Facility (MPCDF) for access to its computing resources.
KW, FFD, and RS have been supported by the German Science Foundation (DFG, project Sp 345/24-1).
This material is based upon work supported by Tamkeen under the NYU Abu Dhabi Research Institute grant CASS. 
The work of TP was supported by the Project No. BR34836926 "Research and monitoring of near-Earth and deep space through the development and scaling of the Kazakhstan optical telescope network", financed by the Aerospace committee of the Ministry of Artificial Intelligence and Digital Development of the Republic of Kazakhstan. 
SL acknowledges the support of the Strategic Priority Research Program of Chinese Academy of Sciences (No.XDB0500203) and the National Natural Science Foundation of China (NSFC 12473017).
SZ acknowledges the support from the National Key Research and Development Program of China (No. 2024YFA1611603) and the Yunnan Key Laboratory of Survey Science (No. 202449CE340002).
RS acknowledges NAOC International Cooperation Office for its support in 2023, 2024, and 2025, and the support by the National Science Foundation of China (NSFC) under grant No. 12473017. This research was supported in part by grant NSF PHY-2309135 to the Kavli Institute for Theoretical Physics (KITP); 
RS gratefully acknowledges hospitality by MPA and Thorsten Naab during frequent visits. 

\end{acknowledgements}

\bibliographystyle{aa}
\bibliography{bibliography}


\appendix

\section{Classification of TDEs} \label{appendix:TDE_class}
For eccentric orbits, if the maximum energy of the debris of the destroyed star is negative, $\epsilon_{\star}+\Delta\epsilon\leq0$, all of the debris after the disruption is bound to the black hole. The condition $\epsilon_{\star}=-\Delta\epsilon$ then gives a critical value of the eccentricity of the star:
\begin{equation}
    \label{eqn:ecrit1}
    e_{\text{crit,1}} = 1 - 2\frac{q^{-1/3}}{\beta}.
\end{equation}
Here, $q$ is the mass ratio of the SMBH and the star, $M_{\rm \text{SMBH}}/m_{\star}$.
If the orbital eccentricity of the star is less than this critical eccentricity, all of its debris will fall into the black hole. This TDE is classified as an eccentric TDE. If the orbital eccentricity of the star is greater than this critical eccentricity but less than 1, then it is expected that more than half of the debris are captured, and the TDE is a marginally eccentric TDE.

A different condition is needed for hyperbolic orbits. If the minimum energy of the debris is positive, $\epsilon_{\star} - \Delta\epsilon \geq 0$, then no debris is bound to the black hole. Then, the condition $\epsilon_{\star}=\Delta\epsilon$ gives another critical value of orbital eccentricity,
\begin{equation}
    \label{eqn:ecrit2}
    e_{\text{crit,2}} = 1 + 2\frac{q^{-1/3}}{\beta}
\end{equation}
above which all of the debris escape from the black hole. If the orbital eccentricity of the star is greater than this critical eccentricity, then the TDE is hyperbolic. If the orbital eccentricity of the star is less than this critical eccentricity but greater than 1, then it is expected that less than half of the debris are captured, and the TDE is a marginally hyperbolic TDE. Parabolic TDEs are when the orbital eccentricity of the star is exactly equal to 1, whose bound mass fraction is expected to be 0.5.

To facilitate a unified classification of TDE types, we introduce a dimensionless parameter, $\chi(e,\beta)$. Because each star has its own critical eccentricities that depend on its specific orbital energy and penetration factor, classifying TDEs directly in terms of these quantities becomes cumbersome. The parameter $\chi(e, \beta)$ collapses these star-dependent critical eccentricities into universal constants, allowing for a clearer and more systematic classification of TDE types. This parameter is defined as
\begin{equation} \label{eqn:chi}
    \chi(e,\,\beta) \equiv \frac{\beta(e-1)}{2} \left( \frac{M_{\rm SMBH}}{m_{\star}} \right)^{1/3}.
\end{equation}
Substituting \hyperref[eqn:ecrit1]{Eq.~\ref*{eqn:ecrit1}} and \hyperref[eqn:ecrit2]{Eq.~\ref*{eqn:ecrit2}} into \hyperref[eqn:chi]{Eq.~\ref*{eqn:chi}} , the critical eccentricities correspond to constant values $\chi_{\rm crit,1} = -1$ and $\chi_{\rm crit,2} = 1$. TDEs can therefore be classified as eccentric ($\chi < -1$), marginally eccentric ($-1 \le \chi < 0$), parabolic ($\chi = 0$), marginally hyperbolic ($0 < \chi \le 1$), and hyperbolic ($\chi > 1$).

The bound mass of the stellar debris in the case of marginally eccentric and marginally hyperbolic TDEs is calculated by integrating the mass fallback rate, $dm/dt$. The mass distribution of the debris depends on the density profile of the star, which is closely related to the stellar mass and age. In this work, we assume, for simplicity, that all stars have the same internal density profile and use their average stellar density. Since the stellar debris obtains the overall energy spread $\simeq 2\Delta\epsilon$, it leads to an estimated distribution $dm/d\epsilon \simeq m_{\star}/2\Delta\epsilon$ \citep{rees88, enk89}. This can also be acquired from a hydrodynamical simulation of a tidal disruption of a main-sequence star with stellar mass $1M_{\odot}$ as a top-hat distribution model \citep{Park_Hayasaki}. Under this assumption, stars are completely destroyed when $\beta \geq 1$.

The lower bound of the integral is the orbital period of the innermost debris,
\begin{equation}
t_{\text{min}} = \frac{\pi}{\sqrt{2}}\frac{GM_{\rm SMBH}}{|\epsilon_{\rm min}|^{3/2}},
\end{equation}
where $\epsilon_{\rm min}=\epsilon_{\star}-\Delta\epsilon$ is the energy of the innermost debris. $t_{\text{min}}$ then can be expressed as 
\begin{equation}
t_{\rm min} = \pi\sqrt{\frac{r_{\star}^3}{2Gm_{\star}}}\left[ q^{-1/3} + \frac{\beta(1-e)}{2}  \right]^{-3/2}.
\end{equation}
The upper bound is infinity because some debris is expected to escape from the black hole. Then, the total bound mass is obtained by integrating $dm/dt$:
\begin{align}
    m_{\text{bound}} 
    &= \int_{t_{\rm min}}^{\infty} \frac{m_{\star}}{2\Delta\epsilon} \frac{1}{3}(2\pi GM_{\rm SMBH})^{2/3}t^{-5/3}\,dt \nonumber \\[12pt]
    &= \frac{m_{\star}}{2} \left[1 + \frac{\beta(1-e)}{2}\left(\frac{M_{\rm SMBH}}{m_{\star}}\right)^{1/3} \right].
\end{align}
This formula also confirms  that the star moving in a parabolic orbit loses half of its mass when it is torn apart.


\section{Unit Conversion For Our Models} \label{appendix:unit}
We summarize the conversion between dimensionless \nbody\ (NB) units and the corresponding physical scaling adopted in this work. The conversions are derived from the initial total cluster mass and length scaling of the simulated NSC model and are provided for reference throughout the paper.

\begin{table}[h]
\centering
{\footnotesize
\begin{tabular}{c|c} 
 \hline \\[-8pt]
    1NB Unit & Physical Scaling \\ [1pt]
 \hline \\[-8pt]
    Time & $5.42\times10^{-2}\,\mathrm{Myr}$ \\
    Mass & $7.59\times10^{4}\,\msun$\\
    Distance & $1.0\,\mathrm{pc}$ \\
 \hline 
\end{tabular}
} \vspace{5pt}
\caption{Unit conversion between N-body units and physical units. The conversion follows the standard Hénon-unit scaling described by \citet{heggie_standardised_1986}.} 
\label{tab:unit}
\end{table}


\section{Virial Equilibrium and Mass Segregation in our models} \label{appendix:additional_figures}
\begin{figure}
    \centering
    \includegraphics[width=1.0\columnwidth]{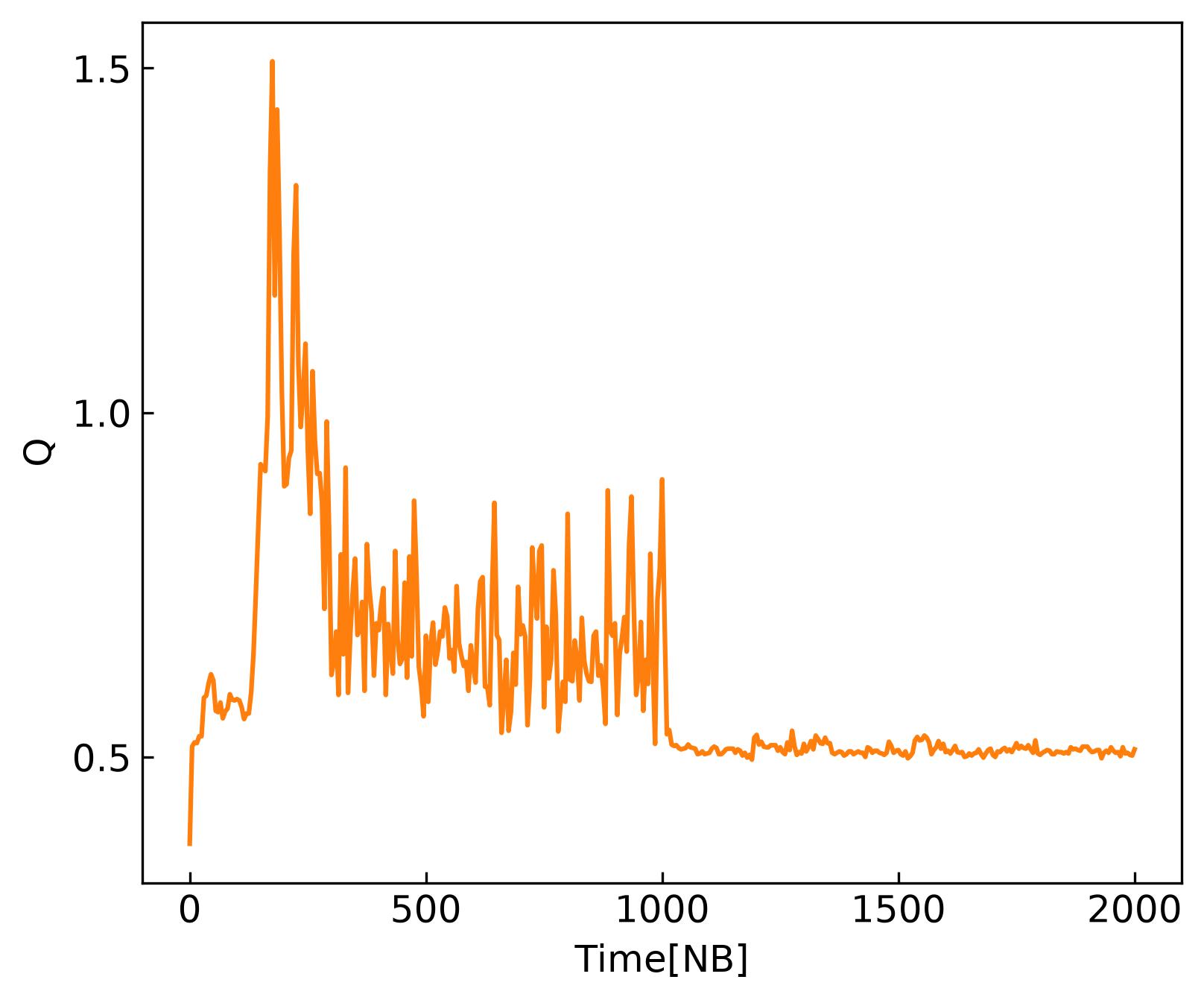}
    \caption{Evolution of the virial ratio $Q$ in the PMA model, with $Q=0.5$ representing virial equilibrium. The cluster completes its initial dynamical adjustment within approximately 1 NB time unit, corresponding to roughly one initial dynamical timescale. Subsequent mass loss occurs due to stellar evolution, which raises $Q$ and produces pronounced fluctuations. At $t\sim1000$ NB, these fluctuations subside and $Q$ stabilizes near 0.5. }
    \label{fig:virial_ratio}
\end{figure}

\autoref{fig:virial_ratio} shows the evolution of the virial ratio $Q$ in the PMA model. The initially sub-virial cluster rapidly adjusts to the combined potential of the NSC and SMBH, reaching $Q\simeq0.5$ within approximately 1 NB time unit, corresponding to roughly one initial dynamical timescale. The subsequent rapid increase in $Q$, followed by an overall decline, is driven mainly by mass loss from stellar evolution, which alters the virial balance of the system and produces strong fluctuations in $Q$. As the cluster dynamically readjusts, these fluctuations subside, and $Q$ stabilizes near 0.5 by $t\sim1000$ NB. Only the PMA model is shown because the FMA model exhibits similar evolution.

\begin{figure}
    \centering
    \includegraphics[width=1.0\columnwidth]{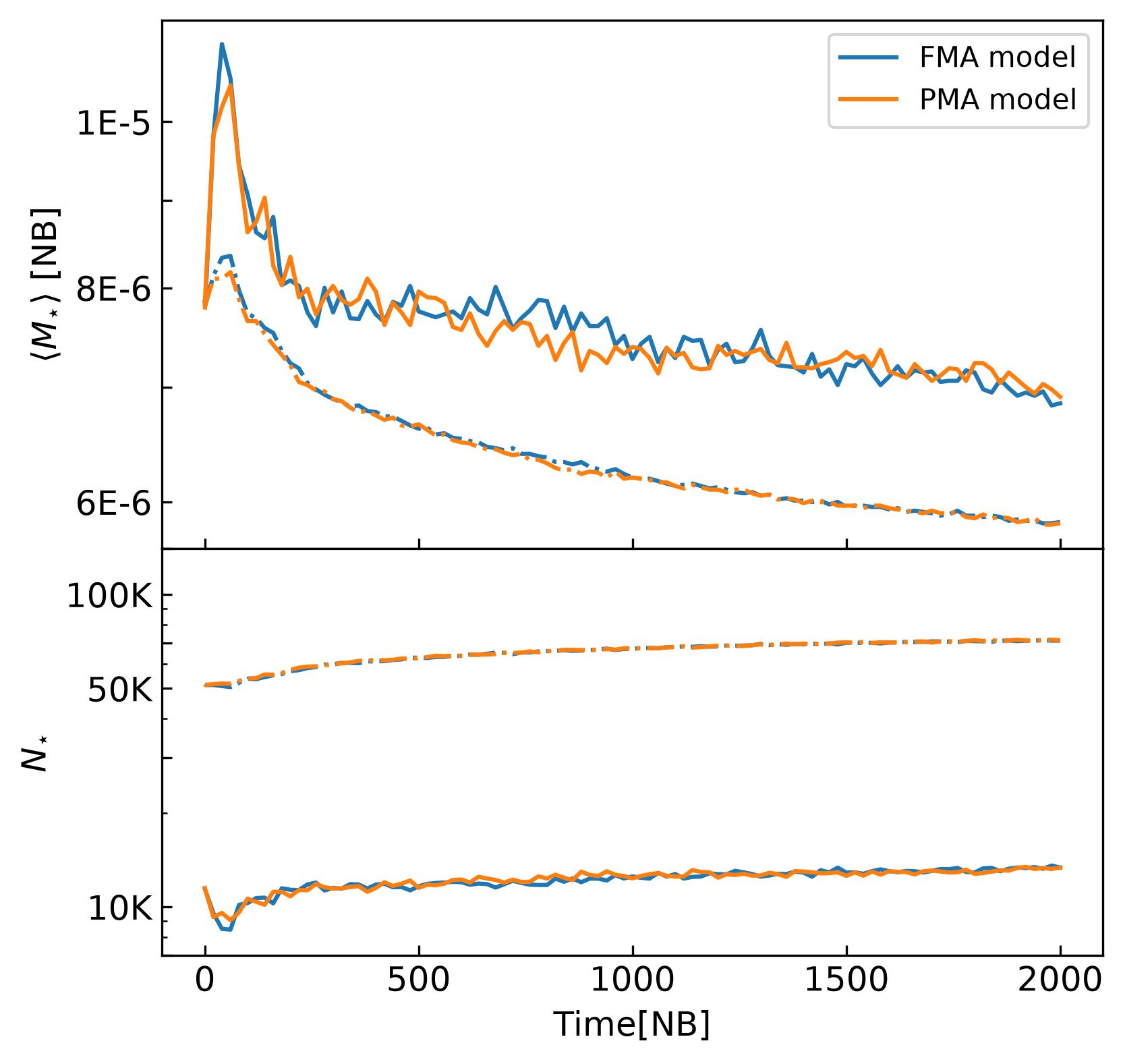}
    \caption{Evolution of the average stellar mass (top) and the number of particles (bottom) within different Lagrangian shells. The solid lines represent the inner shell enclosing 1--10\% of the initial cluster mass, while the dash-dotted lines show the shell enclosing 10--50\%.}
    \label{fig:avmass}
\end{figure}

\autoref{fig:avmass} shows the evolution of the average stellar mass and particle number  in different Lagrangian shells. The gradual decrease in average stellar mass reflects the mass loss driven by stellar evolution, which contributes to the long-term expansion of the cluster. The initial increase in the average mass of the inner shell (1--10\%) is triggered by the removal of stars through TDEs, temporarily shifting the local mass distribution toward higher average values. The corresponding decline in particle number further illustrates the depletion of stars in the central region.

\section{Profiling of Nbody6++GPU} \label{appendix:profiling}

\begin{figure}[h!]
\begin{center}
\includegraphics[width=0.9\columnwidth]{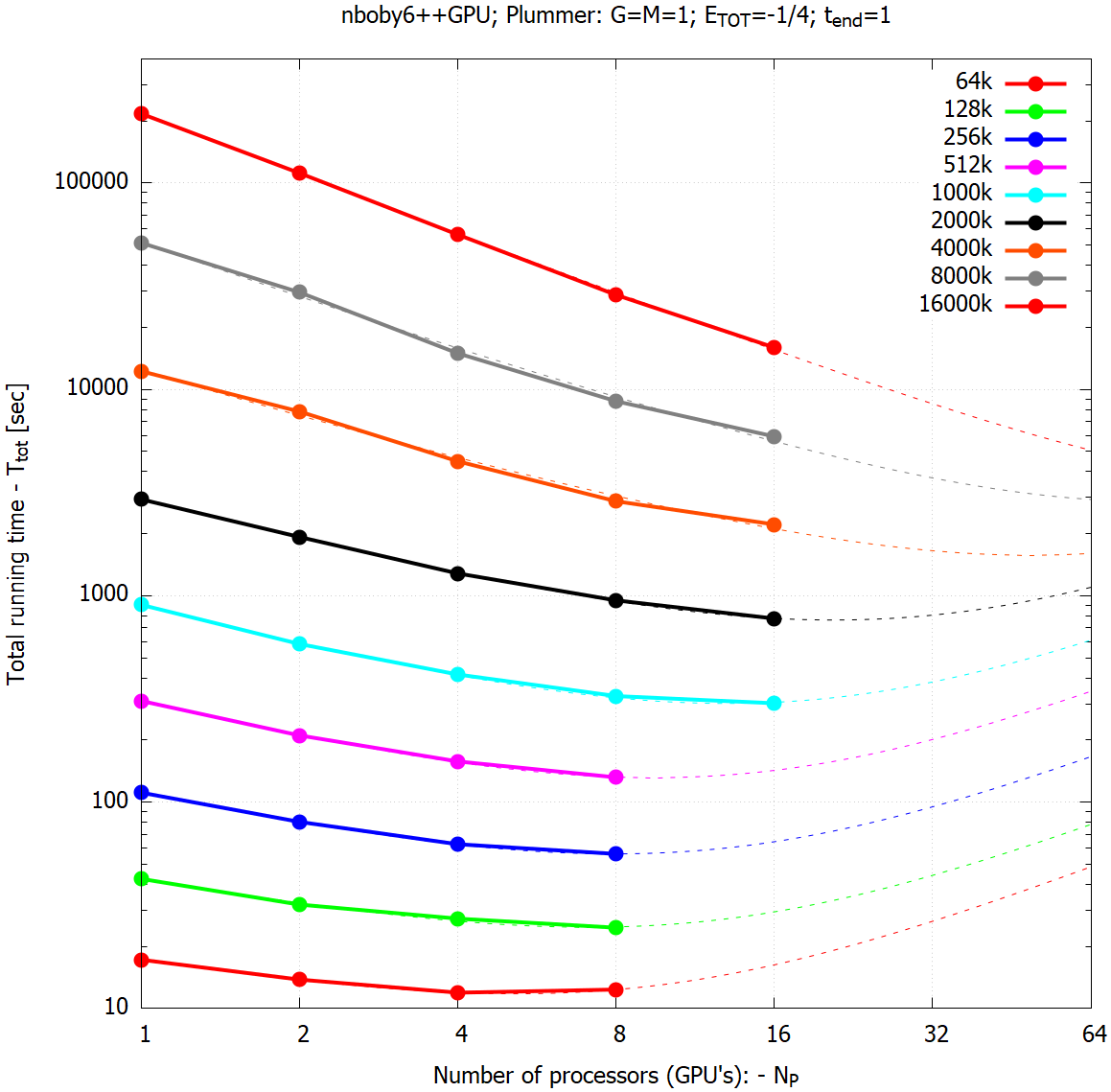}
\includegraphics[width=0.9\columnwidth]{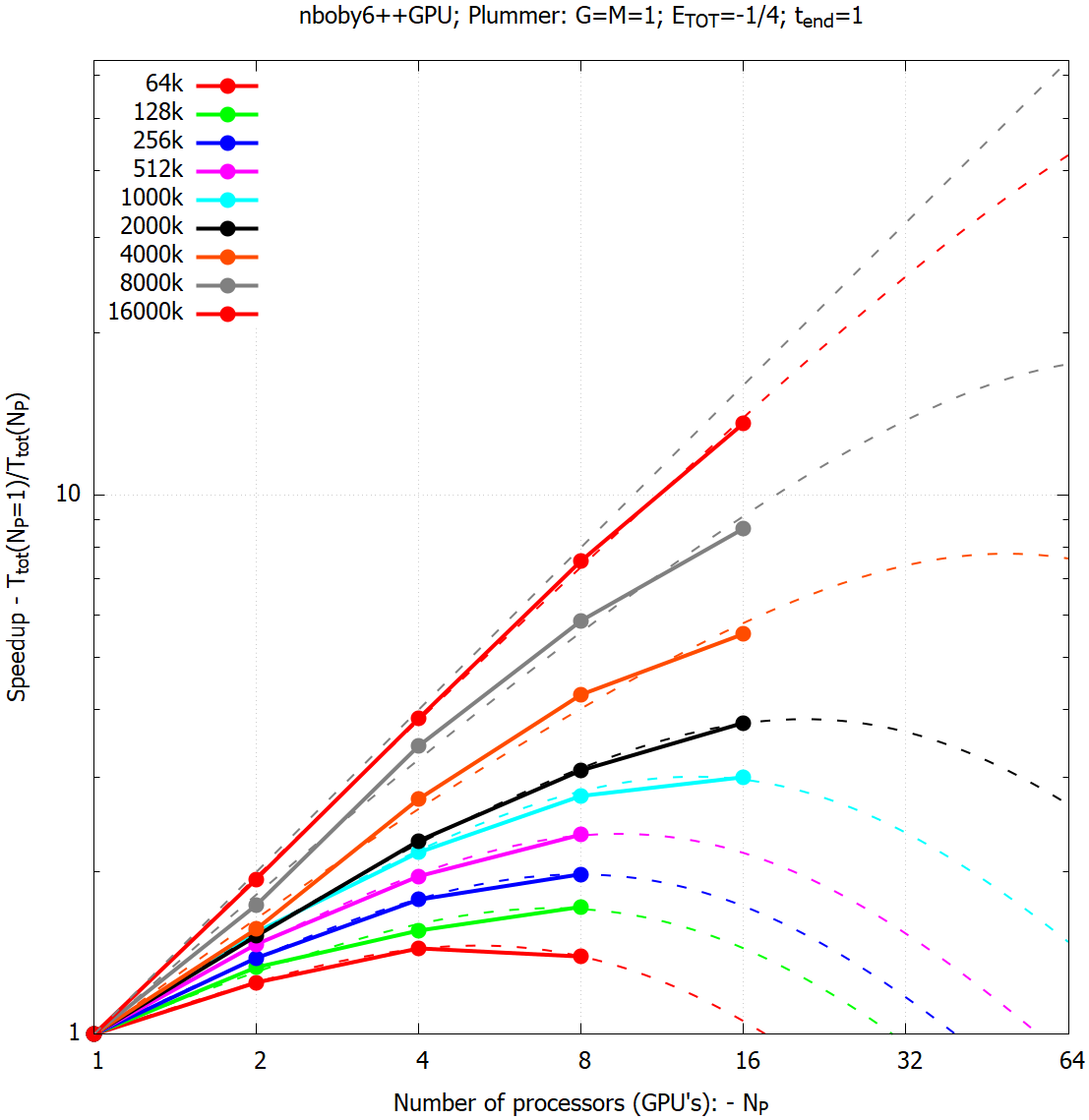}
 \caption{Benchmark results and extrapolated scaling for \textsc{Nbody6++GPU}, initial Plummer model, on the raven cluster at MPCDF, see main text. \textbf{Top:} Total time for one \textsc{Nbody} model unit in secs; \textbf{Bottom:} Speed-Up compared to using one GPU only. In both cases different curves for particle numbers from 64k to 16m. Ideal Speedup is the diagonal dashed lines, other dashed lines extrapolations from the timing model.
\label{fig:nbodyscaling}
}
\end{center}
\end{figure}

We show benchmark results (wall clock time and speedup) in Fig.~\ref{fig:nbodyscaling} obtained on the raven cluster at MPCDF. The results show the potential of our code for future simulations with much larger particle numbers than used in the current paper for pilot simulations. Dots denote measured times of profiling runs (connected by straight lines), while the dashed lines give a fit obtained from our timing model published in \cite{Huang_2016}, with numerical parameters fitted to the raven hardware. 

\label{lastpage}
\end{document}